\documentclass{JHEP3}

\usepackage{amsmath, graphicx, amssymb, cite, afterpage}

\preprint{IPPP/04/84\\DCPT/04/168\\hep-th/0412206}

\author{S. A. Abel and B. W. Schofield\\
        Institute for Particle Physics Phenomenology and Department of Mathematical Sciences, University of Durham, Durham, DH1 3LE, UK\\
        E-mail: \email{s.a.abel@durham.ac.uk}, \email{b.w.schofield@durham.ac.uk}}

\title{One-loop Yukawas on Intersecting Branes}

\abstract{We calculate Yukawa interactions at one-loop on intersecting D$6$ branes. We demonstrate the non-renormalization theorem in
supersymmetric configurations, and show how Yukawa $\beta$-functions may be extracted. In addition to the usual logarithmic running, we
find the power-law dependence on the infra-red cut-off associated with Kaluza-Klein modes. Our results may also be used to evaluate
coupling renormalization in non-supersymmetric cases.}


\begin{document}

\newcommand{\me}{\mathrm{e}}
\newcommand{\mi}{\mathrm{i}}
\newcommand{\sdif}[1]{\!\mathrm{d}#1\;}

\section{Introduction\label{sec:intro}}
Open string models based on intersecting D-branes (see \cite{Uranga:2003pz,Kiritsis:2003mc,Lust:2004ks,Blumenhagen:2004vz} for recent
 reviews) have stimulated new approaches in a number of areas of unification physics. They have also proven to be a useful laboratory for
 testing ideas initially presented in extra-dimensional field theories, without any of the concomitant renormalizability or finiteness
 problems.

Chief amongst these, for the purposes of this paper, are ideas concerning Yukawa couplings and their possible hierarchies. One striking
feature of intersecting branes is their natural replication of families at different intersections, which naturally leads to the idea that
Yukawa hierarchies have a geometrical origin \cite{Cremades:2003qj,Cremades:2004wa}: small Yukawa couplings can arise if different
families are located at different intersections, and the couplings can be exponentially suppressed by world-sheet instantons whose actions
are the areas of the `Yukawa triangles'. This type of picture has its equivalent in closed string orbifold models where the couplings
between twisted states at different fixed points are similarly suppressed \cite{Casas:1991ac,Abel:2002ih,Kobayashi:2003vi,Ko:2004ic}.
There has been significant interest in the phenomenological implications of such a set-up and many extra-dimensional ideas (for example,
the contribution of Kaluza-Klein states to flavour changing \cite{Abel:2003fk}) found their natural realization here. On a more formal
level, Yukawa couplings have important applications in understanding brane recombination processes \cite{Erdmenger:2003kn}.

Going beyond tree-level, there are ideas about Yukawa couplings that have not yet been addressed in a string theory context. One
that will concern us here is power law running. Any model with extra dimensions enjoys the possibility of greatly accelerated
renormalization group running of gauge and Yukawa couplings due to the contribution of Kaluza-Klein states. This effect was
originally suggested in the context of large extra dimensions in refs. \cite{Dienes:1998vg,Dienes:1998vh}, based on the ultra-violet
cut-off dependence of couplings at one loop in field theory with extra dimensions. It has since been explored for gauge couplings in
extra-dimensional field theory (despite its non-renormalizability) by calculating one-loop vacuum polarization diagrams off-shell
and computing their energy-scale dependence in various regularization schemes \cite{Ghilencea:2001bv,Ghilencea:2004sq}.  A genuine
energy-scale dependence has been established and agrees (up to subdominant scheme dependent pieces) with the IR-cut-off dependence
in various string compactifications \cite{Taylor:1988vt,Antoniadis:1999ge}.

There are a number of reasons why a similarly direct comparison of power law running has not yet been done for the Yukawa couplings, most
notably the fact that the tree-level Yukawa couplings are set (at least in the intersecting brane picture) by non-perturbative classical
world sheet instantons. There is no easy prescription for inserting these non-perturbative tree-level couplings into one-loop diagrams
other than simply truncating to the extra dimensional field theory, which would be begging the question. If we want to derive power law
running of Yukawas from intersecting brane configurations, we have to make sure that the \emph{one-loop} classical instantons give the
expected factors of \emph{tree-level} Yukawa couplings in the field theory limits of the string diagram. An additional complicating factor
is the technical difficulty of calculating the one-loop correlation functions that are required to describe interactions between fields
living at intersections.

\FIGURE[t]{\includegraphics{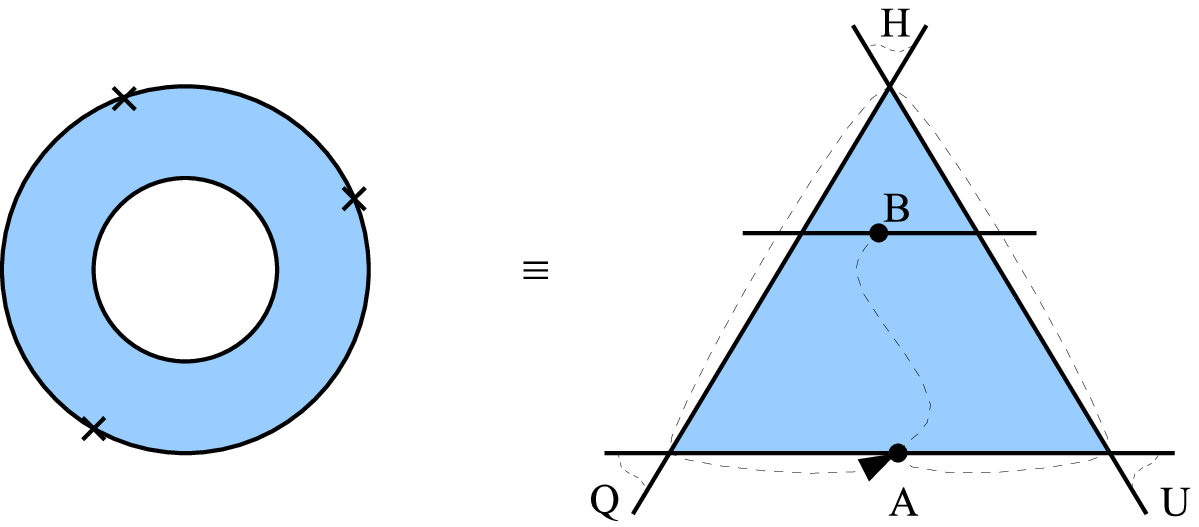}
\caption{The annulus diagram corresponds to taking an open string stretched between two branes as shown, and moving one end around the
Yukawa triangle.\label{fig:annulus}}} So, with this aim in mind, we present the calculation of Yukawa couplings at one-loop on
intersecting branes. (A more general motivation for the study in the present paper is simply that we would like to know how to do
perturbation theory at one-loop on intersecting branes, anyway.) Our main focus will be the annulus diagram with no orbifold twists, shown
in figures \ref{fig:annulus} and \ref{fig:limits}.

We will for concreteness consider D$6$ branes (although the techniques can easily be extended to other configurations) intersecting at
angles in a factorizable torus $T_{2}\times T_{2}\times T_{2}$ whose sub-tori may be tilted and may contain orientifold planes. A general
$\mathcal{N}=1$ set-up will usually involve orbifolds and orientifolds so that there will also be twisted diagrams if the D-branes go
through orbifold fixed points, and there will be M\"obius strip diagrams as well. All of the techniques we are going to describe can be
used for those diagrams as well as the untwisted annulus. However, one can get all the extra information required about these additional
diagrams by factorizing on the one-loop partition function, and so it is possible to present results quite generally in terms of the
latter. The nett effect of the other diagrams is, as one would expect, simply to add new twisted sectors or to project out states in the
spectrum. Consequently, we will focus on the untwisted diagram, but our results will are easily converted to the twisted case.

The first figure spells out the physical principle of the calculation, discussed in ref. \cite{Abel:2003yh}. This is to take a string
stretched between two branes as shown and keep one end (B) fixed on a particular brane, whilst the opposing end (A) sweeps out a Yukawa
triangle. Quark and Higgs states are deposited at each vertex of the triangle as the endpoint A switches from one brane to the next. The
corresponding worldsheet diagram is then the annulus with two fermion and two boson vertex operator insertions on the boundary. There is
no constraint on the relative positioning of the B brane (although the usual rule that the action goes as the square of the brane
separation will continue to be obeyed), and it may be one of the other three branes (in fact, all renormalization diagrams in the
effective field theory will be of this type, and diagrams involving four branes generically separated correspond to string scale masses
circulating in the loop).

In the second figure we have mapped the annulus world-sheet to the rectangular domain shown, which has width $\frac{1}{2}$ and height $\mi
t$. The `branch-cuts' are there to indicate that many correlation functions (for example those involving $\partial X$) will get a phase as
they go round the vertex. This takes account of the change in angle of the allowed motion of the string end-point A as it switches from
one brane to another at the intersection. The regions between vertices then correspond to separate branes (with corresponding phases of
$\partial X$), so the region with no branch cut on the left is the \emph{same brane as} (or at least parallel to) the brane on the
right. The technology of mapping a triangle with a hole in it to a fundamental domain such as this is known as Schwarz-Christoffel
mapping. It was discussed in ref. \cite{Abel:2003yx} for the tree level case, so we will not dwell on it here except to say that we expect
the correlation functions to involve products of elliptic functions in the one loop case. \textbf{}Note that it is possible to have branes
on the right that are not parallel to any on the left in which case one has branch-cuts up the entire height of the fundamental domain. We
will not consider this possibility for reasons that will become clear shortly.

The overriding goal of this calculation will be to show how to recover beta functions for the Yukawa couplings in the field theory limit.
We will therefore mainly be interested in the limit of large $t$ (where $t$ is the ratio of the annulus length to its width, and plays the
role of the Schwinger time) and in particular the dependence of the results on the IR (large-$t$) cut-off. In this limit the beta
functions are dominated by the various field theory limits in which one or more of the vertices are pinched together. The relevant
diagrams for discussing beta functions will be those that factor on a Yukawa coupling times field renormalization diagram. There are four
different limiting cases shown in figure \ref{fig:limits}, which we will refer to as limits 0 to 4. Limit 0 is the partition function
factorization limit where all the vertices come together, and the string diagram factorizes into the product of a one-loop annulus diagram
and the tree-level Yukawa coupling.

Adjacent to the other diagrams are their nearest field theory equivalents.  Concentrating on the quantum part of the amplitude for a
moment, limit 1 (non-degenerate vertices) is a coupling renormalization diagram and so in supersymmetric theories this `limit' should give
zero. In $\mathcal{N}=0$ models such diagrams will be non-zero and will represent an actual coupling renormalization. (However, the
$t\rightarrow0$ limit would yield UV divergences in these cases due to non-vanishing tadpoles in the closed string channel, indicating a
non-trivial background.)  One of our tasks therefore will be to show the vanishing of this contribution in supersymmetric configurations
due to a $\vartheta$-function identity.  This is the stringy version of the non-renormalization theorem. The only opportunity to obtain
non-zero contributions is therefore when there is a pole, corresponding to limits 2 and 3.

One diagram that we will not consider is the the one which would be a Yukawa renormalization with three intersection (twisted) fields in
the loop. It is this case which corresponds to the diagram with a non-parallel fourth brane and branch-cut all the way up. Calculating
them would involve a significant complication, but since they can only be relevant for non-supersymmetric theories anyway, we feel
justified in neglecting them.

\afterpage{\clearpage}
\FIGURE{
\begin{tabular}{cccc}
\parbox[t][10mm][t]{10mm}{\bf 0.}&
\parbox[c][45mm][c]{35mm}{\includegraphics{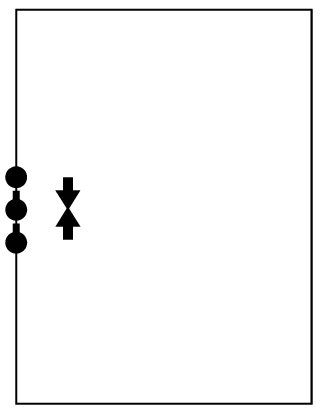}}&
\parbox[c][\height][c]{7mm}{$=$}&
\parbox[c][\height][c]{34mm}{\includegraphics{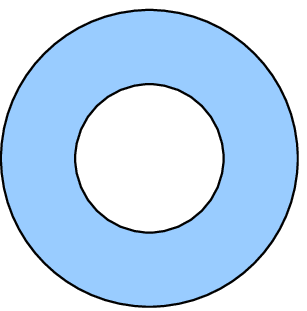}}\parbox[c][\height][c]{13mm}{$\quad\times\quad$}\parbox[c][\height][c]{30mm}{\scalebox{0.7}{\includegraphics{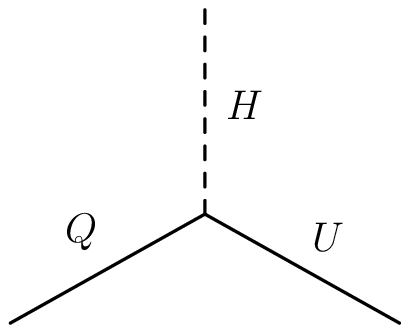}}}\tabularnewline
\parbox[t][10mm][t]{10mm}{\bf 1.}&
\parbox[c][45mm][c]{35mm}{\includegraphics{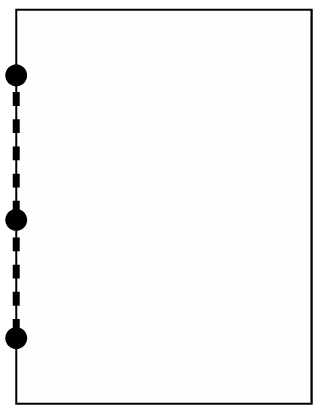}}&
\parbox[c][\height][c]{7mm}{$=$}&
\parbox[c][\height][c]{40mm}{\scalebox{0.7}{\includegraphics{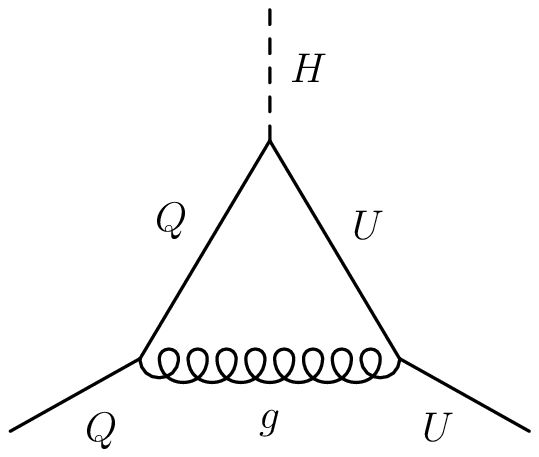}}}\tabularnewline
\parbox[t][10mm][t]{10mm}{\bf 2.}&
\parbox[c][45mm][c]{35mm}{\includegraphics{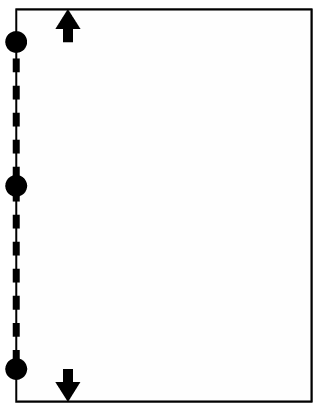}}&
\parbox[c][\height][c]{7mm}{$=$}&
\parbox[c][\height][c]{25mm}{\scalebox{0.7}{\includegraphics{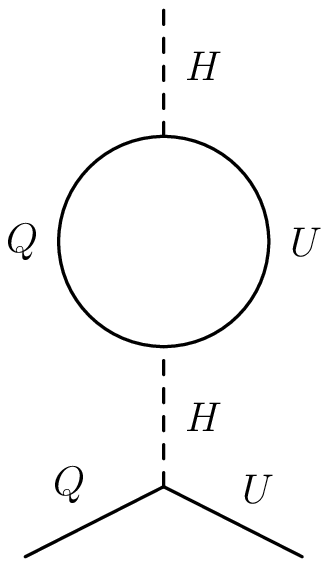}}}\parbox[c][\height][c]{13mm}{$\quad+\quad$}\parbox[c][\height][c]{25mm}{\scalebox{0.7}{\includegraphics{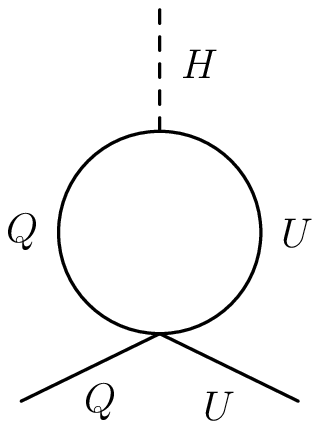}}}\tabularnewline
\parbox[t][10mm][t]{10mm}{\bf 3.}&
\parbox[c][45mm][c]{35mm}{\includegraphics{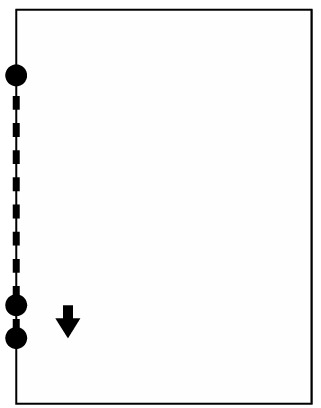}}&
\parbox[c][\height][c]{7mm}{$=$}&
\parbox[c][\height][c]{25mm}{\scalebox{0.7}{\includegraphics{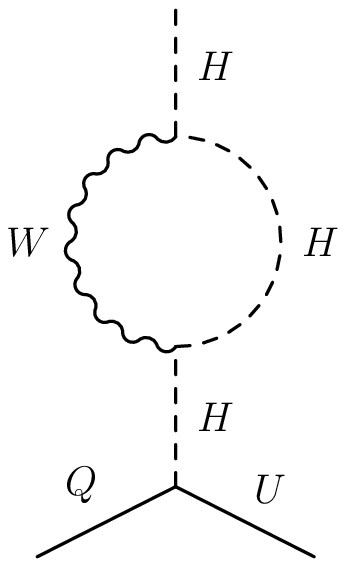}}}\tabularnewline
\end{tabular}
\caption{Field-theory diagrams from various limits on the worldsheet.\label{fig:limits}} }

Our expectations for the classical worldsheet-instanton contributions to the amplitudes are based on the quantum part. Field theoretically
the closest diagram to limit 1 would be proportional to $\mathcal{Y}g^{2}$ where $\mathcal{Y}$ is the Yukawa coupling and $g$ the relevant
gauge coupling of the gauge field in the loop. The Yukawa coupling at tree level is determined by the worldsheet instanton whose action is
simply the sum of the projected triangle areas swept out by the worldsheet in the three sub-tori,
$S_{\text{tree}}=\sum_{i=1}^{3}\left(\text{Area}\right)_{i}$ (the tree-level Yukawa is proportional to $e^{-S_{\text{tree}}}$).  The same
sum of triangle areas should appear in the one-loop action in this limit.

Limit 2 gives the field theory diagram corresponding to a Yukawa contribution with a bubble on one of the legs, which has two twisted
states running in the loop. We can see this heuristically by noting that the imaginary direction on the annulus represents the loop, and
that the branch-cut free part of the annulus is pinched in this limit. These contributions should be proportional to
$\mathcal{Y}\mathcal{Y}^{\dagger}\mathcal{Y}$.  This means that the one-loop classical action should yield $S_{\text{one-loop}}=\sum
S_{\text{tree}}$ in this limit, where the sum is over Yukawas appearing in the field theory diagram. The diagram should also have a
$1/k^{2}$ pole that requires the $k^{2}$ term in the quantum prefactor; this is simply the contribution to the Yukawa beta function coming
from field renormalization.  As all states are twisted, there should be no Kaluza-Klein (KK) modes in this limit.

Limit 3 gives the field theory diagram with one twisted state and one untwisted (gauge) state in the loop. These contributions should
still be proportional to $\mathcal{Y}g^{2}$. Note that we should get contributions from all the KK modes of the gauge field, generating
power-law running of the Yukawa coupling in this limit. This sum over KK contributions comes from a Poisson resummation of the classical
action contribution to the amplitude.

The paper is organized as follows. In the next section we set up the string theory calculation, discussing vertex operators and charge
conservation, and extract the general form of the amplitudes in terms of correlators. We then evaluate all necessary correlators,
including the spin and twist fields, on the annulus. In section \ref{sec:quantum}, we collect the necessary correlators together and
discuss the quantum part of the amplitude, elucidating the emergence of the non-renormalization theorem. Section \ref{sec:classical} is
devoted to deriving the classical instanton action via monodromy conditions. Section \ref{sec:classical-limits} discusses the
factorization of the classical part onto the various limits discussed above, while section \ref{sec:beta} discusses the extraction of the
beta functions and power law running.

\section{Elements of the calculation\label{sec:stringcalc}}

Before we begin, it is perhaps worthwhile to give a brief review of background literature. The supersymmetric non-renormalization theorem
has been demonstrated explicitly for a $\mathbb{Z}_{3}$ orbifold using only untwisted states \cite{Mandal:1987kz}, and more general rules
have also been derived for other $D=4$ models \cite{Lechtenfeld:1989be}.  Also, an extraction of the $\beta$-function for gauge fields was
performed for supersymmetric orbifolds \cite{Minahan:1987ha}.

Key elements in our calculation will be the conservation of the $H$-charge of bosonized states, correlation functions of spin operators
and correlation functions of twist operators. $H$-charge conservation and consequent selection rules are discussed for orbifold models in
\cite{Cvetic:1987ac,Cvetic:1987fw,Cvetic:1987qx,Font:1988mm,Font:1988nc,Font:1988tp,Kobayashi:1995py}.  Spin field operators for closed
strings at one loop have been given explicitly for flat backgrounds \cite{Atick:1986es,Atick:1986rs,Atick:1987ns}, and implicitly for
orbifold backgrounds \cite{Atick:1987gy}; we give an explicit result and generalize to open strings in section \ref{sub:corr_spin}.
Similarly, twist field correlators are known at one loop for closed strings \cite{Atick:1988kd}, which we generalize to open strings in
section \ref{sub:corr_twist}. We also examine correlation functions of excited twist operators at one loop, which have been discussed for
the orbifold case in \cite{Dixon:1987qv,Bailin:1993km,Bailin:1993ks}.

\FIGURE[t]{\parbox{\hsize}{\begin{center}
\includegraphics{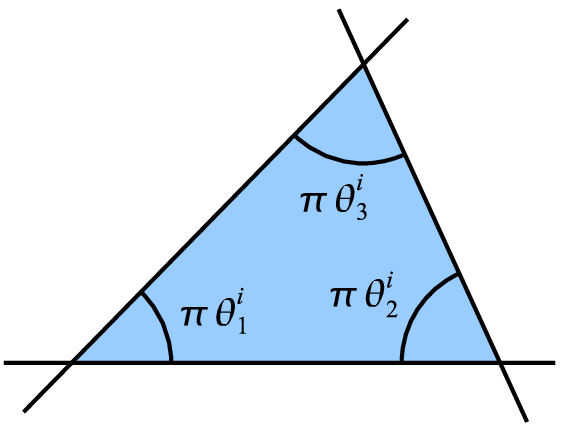}
\caption{We work with internal angles, and always stick to the notation where
subscripts on angles label intersections, and superscripts label tori.\label{fig:triangle-angles}}
\end{center}}}
We will work with type IIA theory compactified on a factorizable $T^{6}$, with D$6$-branes at angles wrapping in the compact space. In
each sub-torus, the intersection of the branes makes a triangle. The rotations which take us from one brane to another are taken as shown:
hence, $\theta_{1}^{i}+\theta_{2}^{i}+\theta_{3}^{i}=1$ in each sub-torus.  Unless displayed explicitly, we set
$\alpha^{\prime}=\frac{1}{2}$.

\subsection{Vertex operators and $H$-charge\label{sub:vertex_and_pcos}}

At any given intersection, an analysis of the mode expansions of strings stretched between the branes~\cite{Aldazabal:2000dg} tells us
that we obtain from the R sector a massless fermion. Introducing a set $H_{i}$ of bosonic fields on the worldsheet, where $i$ labels pairs
of complex dimensions, the most canonical vertex operator for this state is
\cite{Berkooz:1996km,Cvetic:2003ch,Klebanov:2003my}\begin{equation}
\mathcal{V}_{-\frac{1}{2}}\left(u,k,z\right)=\me^{-\frac{1}{2}\phi}uS^{\pm}\me^{\mi k\cdot X}\me^{\mi q\cdot
H}\Sigma\left(z\right)\quad\text{with}\quad
q=\left(\tfrac{1}{2}-\theta^{1},\,\tfrac{1}{2}-\theta^{2},\,\tfrac{1}{2}-\theta^{3}\right)\,,\label{eq:vf}\end{equation} where $u$ is the
four-dimensional polarization, $S^{\pm}=\me^{\pm\frac{1}{2}\mi\left(H_{1}+H_{2}\right)}$ is a four-dimensional spin field, $\me^{\mi
q\cdot H}$ is the spin field for the internal co-ordinates, and $\Sigma=\sigma^{1}\sigma^{2}\sigma^{3}$ are bosonic twist fields
\cite{Dixon:1987qv,Hamidi:1986vh}. Our convention for angles means that the operators $\sigma$ correspond to `anti-twist' fields,
sometimes denoted $\sigma_{-}$ or $\sigma_{1-\theta}$ in the literature. The plane wave portion of the vertex operators, $\me^{\mi k\cdot
X}$, is present only in the non-compact space -- $k^{\mu}$ is a four-vector.  We have suppressed factors of the string coupling $g_{O}$
and Chan-Paton factors in this expression, and will do so throughout the rest of this work.

A vertex operator of the opposite chirality may also be created, representing strings stretched with the opposite orientation, but as our
final result ought to be comparable to a quark Yukawa term, in which both fermions have the same chirality, we will not need it.

In the NS sector, the scalars surviving the GSO projection are generally massive~\cite{Aldazabal:2000dg},
\begin{center}\begin{tabular}{ccc}
State&
$q$&
$m^{2}$\vspace{0.5ex}\tabularnewline
\hline
&
\vspace{-1.5ex}&
\tabularnewline
$\Psi_{\frac{1}{2}-\theta^{1}}\left|0\right\rangle $&
$\left(-\theta^{1},1-\theta^{2},1-\theta^{3}\right)$&
$\theta^{1}-\theta^{2}-\theta^{3}+1$\vspace{1ex}\tabularnewline
$\Psi_{\frac{1}{2}-\theta^{2}}\left|0\right\rangle $&
$\left(1-\theta^{1},-\theta^{2},1-\theta^{3}\right)$&
$-\theta^{1}+\theta^{2}-\theta^{3}+1$\vspace{1ex}\tabularnewline
$\Psi_{\frac{1}{2}-\theta^{3}}\left|0\right\rangle $&
$\left(1-\theta^{1},1-\theta^{2},-\theta^{3}\right)$&
$-\theta^{1}-\theta^{2}+\theta^{3}+1$\vspace{1ex}\tabularnewline
$\Psi_{\frac{1}{2}-\theta^{1}}\Psi_{\frac{1}{2}-\theta^{2}}\Psi_{\frac{1}{2}-\theta^{3}}\left|0\right\rangle $&
$\left(-\theta^{1},-\theta^{2},-\theta^{3}\right)$&
$\theta^{1}+\theta^{2}+\theta^{3}-1$\vspace{1ex}\tabularnewline
\end{tabular}\end{center}

With a suitable choice of angles, we may make any one (but only one) of these states massless, in which case it becomes the
$\mathcal{N}=1$ superpartner of the massless fermion in the R sector. The other three states are then massive, and so must be
superpartners of other heavy fermions.

From factorization of the tree level four-point function~\cite{Abel:2003yx}, we identify the Higgs as the state
$\Psi_{\frac{1}{2}-\theta^{1}}\Psi_{\frac{1}{2}-\theta^{2}}\Psi_{\frac{1}{2}-\theta^{3}}\left|0\right\rangle $.  The $\mathcal{N}=1$
supersymmetry condition on our set-up is then\begin{equation} \theta^{1}+\theta^{2}+\theta^{3}=1\,,\label{eq:n1susy}\end{equation} and the
most canonical vertex operator is\begin{equation} \mathcal{V}_{-1}\left(k,z\right)=\me^{-\phi}\me^{\mi k\cdot X}\me^{\mi q\cdot
H}\Sigma\left(z\right)\quad\text{with}\quad q=\left(-\theta^{1},-\theta^{2},-\theta^{3}\right)\,.\label{eq:vb}\end{equation} Again, a
conjugate vertex operator exists but is not of relevance here. One may verify that (\ref{eq:vb}) correctly has unit conformal weight,
provided that $k^{2}=-m^{2}$.

Another way to see that that we have identified the correct state as the Higgs is to note that since the $H$ are a set of free bosonic
fields, non-vanishing terms in $\mathcal{A}$ must obey $H$-charge (momentum) conservation: \begin{equation}
\sum_{i=1}^{3}q_{i}^{j}=0\label{eq:hcharge}\end{equation} in each complex dimension $j$. If we were interested in calculating the
\emph{tree-level} Yukawa coupling,\begin{equation} \mathcal{Y}=\left\langle
\mathcal{V}_{-\frac{1}{2}}\mathcal{V}_{-\frac{1}{2}}\mathcal{V}_{-1}\right\rangle \,,\label{eq:amp_tree}\end{equation} we would see that
$H$-charge conservation is correctly obeyed by the vertex operators given: in an external dimension, $\pm\frac{1}{2}\mp\frac{1}{2}=0$, and
in an internal dimension $\frac{1}{2}-\theta_{1}+\frac{1}{2}-\theta_{2}-\theta_{3}=0$.

\subsection{One-loop amplitudes and picture-changing}

On the annulus, the appropriate string scattering amplitude is given by finding the correlation function of two fermions plus one boson,
then integrating over all possible configurations of vertex operators plus the modular parameter $t$ of the annulus:\begin{equation}
\mathcal{A}=\int_{0}^{\infty}\frac{\mathrm{d}t}{t}f\left(t\right)\int\sdif{z_{1}\mathrm{d}z_{2}\mathrm{d}z_{3}}A\left(z_{1},z_{2},z_{3}\right)\,.\label{eq:amp}\end{equation}
$f\left(t\right)$ is an overall normalization which will be determined later by coalescing the three vertex operators (limit 0 of figure
\ref{fig:limits}). For now, we focus on the computation of $A$.

At tree-level, the amplitude (\ref{eq:amp_tree}) has an overall $\phi$-charge of $-\frac{1}{2}-\frac{1}{2}-1=-2$. On the annulus,
$\mathcal{A}$ must have an overall $\phi$-charge of zero if it is to be non-anomalous, and we have to perform `picture-changing'
operations~\cite{Friedan:1986ge} on our vertex operators:\begin{equation} \mathcal{V}_{i+1}\left(z\right)=\lim_{w\rightarrow
z}\me^{\phi}T_{F}\left(w\right)\mathcal{V}_{i}\left(z\right)\,.\label{eq:pco}\end{equation} Here, $T_{F}$ is the generator of worldsheet
supersymmetry,\begin{equation} T_{F}=\frac{1}{2}(\partial\overline{X}^{i}\psi_{i}+\partial
X^{i}\overline{\psi}_{i})\,,\label{eq:susygen}\end{equation} with $\psi\sim\me^{\mi H}$ and $\overline{\psi}\sim\me^{-\mi H}$ bosonized
fermions in each complex pair of dimensions,\begin{equation} X^{i}=X^{\mu=2i}+\mi X^{\mu=2i+1}\qquad\qquad\overline{X}^{i}=X^{\mu=2i}-\mi
X^{\mu=2i+1}\,.\label{eq:complex}\end{equation} To obtain the correct $\phi$-charge, we need to insert two picture-changing operators
somewhere in our amplitude. It will be convenient to apply one operator to one of the fermions, and one to the boson, so that the relevant
correlation function is\begin{equation} \mathcal{A}=\left\langle
\mathcal{V}_{-\frac{1}{2}}\left(\overline{u}_{1},k_{1},z_{1}\right)\lim_{w_{2}\rightarrow
z_{2}}\me^{\phi}T_{F}\left(w_{2}\right)\mathcal{V}_{-\frac{1}{2}}\left(u_{2},k_{2},z_{2}\right)\lim_{w_{3}\rightarrow
z_{3}}\me^{\phi}T_{F}\left(w_{3}\right)\mathcal{V}_{-1}\left(u_{3},k_{3},z_{3}\right)\right\rangle \,.\label{eq:corr}\end{equation}

Oour picture-changing operators must be inserted in such a way as not to affect the overall $H$-charge, meaning that for each complex
dimension, we should make one $\partial\overline{X}\psi$ and one $\partial X\overline{\psi}$ insertion. In other words, only terms of the
form\begin{equation}
\mathcal{V}_{-\frac{1}{2}}\left(z_{1}\right)\partial\overline{X}^{i}\psi^{i}\left(w_{2}\right)\mathcal{V}_{-\frac{1}{2}}\left(z_{2}\right)\partial
X^{i}\overline{\psi}^{i}\left(w_{3}\right)\mathcal{V}_{-1}\left(z_{3}\right)\label{eq:h_allowed_1}\end{equation} and\begin{equation}
\mathcal{V}_{-\frac{1}{2}}\left(z_{1}\right)\partial
X^{i}\overline{\psi}^{i}\left(w_{2}\right)\mathcal{V}_{-\frac{1}{2}}\left(z_{2}\right)\partial\overline{X}^{i}\psi^{i}\left(w_{3}\right)\mathcal{V}_{-1}\left(z_{3}\right)\label{eq:h_allowed_2}\end{equation}
may contribute to $A$ (no summation over $i$ is implied).

We may see explicitly the effect of the picture-changing operators on the vertex operators (\ref{eq:vf}) and (\ref{eq:vb}) by using the
OPEs\begin{eqnarray} \me^{iaH}\left(w\right)\me^{ibH}\left(z\right) & \sim &
\left(w-z\right)^{ab}\me^{i\left(a+b\right)H}\left(z\right)\nonumber \\ \partial X\left(w\right)\me^{ik\cdot X}\left(z\right) & \sim &
\frac{\mi k}{w-z}\me^{ik\cdot X}\left(z\right)\nonumber \\ \partial\overline{X}\left(w\right)\me^{ik\cdot X}\left(z\right) & \sim &
\frac{\mi\overline{k}}{w-z}\me^{ik\cdot X}\left(z\right)\,,\label{eq:opes_xpsi}\end{eqnarray} where $k^{i}$ is defined in exactly the same
way as $X^{i}$, so that $k\cdot X=\frac{1}{2}\left(k^{i}\overline{X}^{i}+\overline{k}^{i}X^{i}\right)$, and\begin{eqnarray} \partial
X\left(w\right)\sigma\left(z\right) & \sim & \left(w-z\right)^{-\theta}\tau\left(z\right)\nonumber \\
\partial\overline{X}\left(w\right)\sigma\left(z\right) & \sim &
\left(w-z\right)^{-\left(1-\theta\right)}\tau^{\prime}\left(z\right)\,,\label{eq:opes_xsigma}\end{eqnarray} with $\tau$ and
$\tau^{\prime}$ excited twist fields%
\footnote{We have made the replacement $\theta\rightarrow 1-\theta$ with respect to the usual definition of these OPEs, reflecting the fact
that we work with internal angles as shown in figure~\ref{fig:triangle-angles}.}.

Since the vertex operators are segregated into operators which act only in the internal dimensions ($S^{\pm}$, $\me^{\mi k\cdot X}$) and
operators which act only in the external dimensions ($\me^{\mi q\cdot H}$, $\Sigma$), we need to treat internal and external indices
differently.  Beginning with the case where $i$ is an internal index, we have for the fermion,\begin{eqnarray}
\frac{1}{2}\lim_{w\rightarrow z}\me^{\phi}\partial\overline{X}^{i}\psi^{i}\left(w\right)\mathcal{V}_{-\frac{1}{2}}\left(z\right) & = &
\frac{1}{2}\me^{\frac{1}{2}\phi}uS^{\pm}\me^{\mi k\cdot X}\me^{\mi\tilde{q}\cdot H}\tau^{\prime
i}\sigma^{j}\sigma^{k}\left(z\right)\quad\text{with}\quad\tilde{q}^{i}=\tfrac{3}{2}-\theta^{i}\nonumber \\ \frac{1}{2}\lim_{w\rightarrow
z}\me^{\phi}\partial X^{i}\overline{\psi}^{i}\left(w\right)\mathcal{V}_{-\frac{1}{2}}\left(z\right) & = &
\frac{1}{2}\me^{\frac{1}{2}\phi}uS^{\pm}\me^{\mi k\cdot X}\me^{\mi\tilde{q}\cdot
H}\tau^{i}\sigma^{j}\sigma^{k}\left(z\right)\quad\text{with}\quad\tilde{q}^{i}=-\tfrac{1}{2}-\theta^{i}\,,\nonumber \\
\label{eq:pcos_on_vf_int}\end{eqnarray}
again with no summation implied over $i$.

For the boson,\begin{eqnarray} \frac{1}{2}\lim_{w\rightarrow
z}\me^{\phi}\partial\overline{X}^{i}\psi^{i}\left(w\right)\mathcal{V}_{-1}\left(z\right) & = & \frac{1}{2}\me^{\mi k\cdot
X}\me^{\mi\tilde{q}\cdot H}\tau^{\prime i}\sigma^{j}\sigma^{k}\left(z\right)\quad\text{with}\quad\tilde{q}^{i}=1-\theta^{i}\nonumber \\
\frac{1}{2}\lim_{w\rightarrow z}\me^{\phi}\partial X^{i}\overline{\psi}^{i}\left(w\right)\mathcal{V}_{-1}\left(z\right) & = &
0\,.\label{eq:pcos_on_vb_int}\end{eqnarray} Therefore, the internal dimensions contribute only to $\mathcal{A}$ via terms of the form
(\ref{eq:h_allowed_2}). We label these contributions as $A_{1}$:\begin{multline} A_{1}=\frac{1}{4}\overline{u}_{1}u_{2}\left\langle
\me^{-\frac{1}{2}\phi}\left(z_{1}\right)\me^{\frac{1}{2}\phi}\left(z_{2}\right)\right\rangle \left\langle
S^{\mp}\left(z_{1}\right)S^{\pm}\left(z_{2}\right)\right\rangle \left\langle \me^{\mi k_{1}\cdot X}\left(z_{1}\right)\me^{\mi k_{2}\cdot
X}\left(z_{2}\right)\me^{\mi k_{3}\cdot X}\left(z_{3}\right)\right\rangle \\ \sum_{i\ne j\ne k}\left\langle \me^{\mi q_{1}\cdot
H(z_{1})}\me^{\mi\tilde{q}_{2}\cdot H(z_{2})}\me^{\mi\tilde{q}_{3}\cdot H(z_{3})}\right\rangle \left\langle
\sigma^{i}(z_{1})\tau^{i}(z_{2})\tau^{\prime i}(z_{3})\right\rangle \left\langle
\sigma^{j}(z_{1})\sigma^{j}(z_{2})\sigma^{j}(z_{3})\right\rangle \\ \left\langle
\sigma^{k}(z_{1})\sigma^{k}(z_{2})\sigma^{k}(z_{3})\right\rangle \,.\label{eq:a1}\end{multline} The arrangement of picture-changed
$H$-charge in the spin operators here is displayed explicitly in table \ref{tab:H_charge}.  \TABLE[t]{\parbox{\hsize}{\begin{center}
\begin{tabular}{c|ccc}
\begin{picture}(80,40)(0,0)

  \put(30,30){Sub-torus}

  \put(0,40){\line(2,-1){80}}

  \put(0,0){$H$-charge}

\end{picture}&
\raisebox{15pt}{$i$}& \raisebox{15pt}{$j$}& \raisebox{15pt}{$k$}\tabularnewline \hline \rule[3ex]{0ex}{0ex}$q_{1}$&
$\frac{1}{2}-\theta_{1}^{i}$& $\frac{1}{2}-\theta_{1}^{j}$& $\frac{1}{2}-\theta_{1}^{k}$\rule[-2ex]{0ex}{0ex}\tabularnewline
$\tilde{q}_{2}$& $-\frac{1}{2}-\theta_{2}^{i}$& $\frac{1}{2}-\theta_{2}^{j}$&
$\frac{1}{2}-\theta_{2}^{k}$\rule[-2ex]{0ex}{0ex}\tabularnewline $\tilde{q}_{3}$& $1-\theta_{3}^{i}$& $-\theta_{3}^{j}$&
$-\theta_{3}^{k}$\rule[-2ex]{0ex}{0ex}\tabularnewline
\end{tabular}
\caption{$H$-charge assignments in the contribution $\mathcal{A}_{1}$. The
sum in each column is zero, preserving charge conservation in each
sub-torus. \label{tab:H_charge}}
\end{center}}}

Now consider the case when $i$ is an external index. There are two different possibilities for the result of the picture-changing
operation on the fermion operator $\mathcal{V}_{-\frac{1}{2}}$, depending upon whether it contains an
$S^{+}=\me^{+\frac{1}{2}\mi\left(H_{1}+H_{2}\right)}$ or an $S^{-}=\me^{-\frac{1}{2}\mi\left(H_{1}+H_{2}\right)}$ operator.  If
$\mathcal{V}_{-\frac{1}{2}}$ contains an $S^{+}$, then we have\begin{align} \frac{1}{2}\lim_{w\rightarrow
z}\me^{\phi}\partial\overline{X}^{i}\psi^{i}\left(w\right)\mathcal{V}_{-\frac{1}{2}}\left(z\right) & =\frac{1}{2}\mi
k^{i}\me^{\frac{1}{2}\phi}\me^{\mi k\cdot X}S_{\uparrow}^{+}\me^{\mi q\cdot H}\Sigma\left(z\right)\nonumber \\
\frac{1}{2}\lim_{w\rightarrow z}\me^{\phi}\partial X^{i}\overline{\psi}^{i}\left(w\right)\mathcal{V}_{-\frac{1}{2}}\left(z\right) &
=\frac{1}{2}\partial X^{i}\me^{\frac{1}{2}\phi}\me^{\mi k\cdot X}S_{\downarrow}^{+}\me^{\mi q\cdot H}\Sigma\left(z\right)+\text{singular
term}\label{eq:pcos_on_vf_ext_splus}\end{align} where\begin{eqnarray}
S_{\uparrow}^{+}=\me^{\mi\left(\frac{3}{2}H_{i}+\frac{1}{2}H_{j}\right)} & \qquad\text{and}\qquad &
S_{\downarrow}^{+}=\me^{\mi\left(-\frac{1}{2}H_{i}+\frac{1}{2}H_{j}\right)}\,.\label{eq:splus_modified}\end{eqnarray} The singular term
will vanish via the Dirac equation, just as it does in the picture-changing of $\mathcal{V}_{-\frac{1}{2}}$ when no branes are present
\cite{Friedan:1986ge,Polyakov:1995uz}; this vanishing has been obscured by the `helicity basis' used to represent the spin operators.

If $\mathcal{V}_{-\frac{1}{2}}$ contains an $S^{-}$ operator,\begin{align} \frac{1}{2}\lim_{w\rightarrow
z}\me^{\phi}\partial\overline{X}^{i}\psi^{i}\left(w\right)\mathcal{V}_{-\frac{1}{2}}\left(z\right) &
=\frac{1}{2}\partial\overline{X}^{i}\me^{\frac{1}{2}\phi}\me^{\mi k\cdot X}S_{\uparrow}^{-}\me^{\mi q\cdot
H}\Sigma\left(z\right)+\text{singular term}\nonumber \\ \frac{1}{2}\lim_{w\rightarrow z}\me^{\phi}\partial
X^{i}\overline{\psi}^{i}\left(w\right)\mathcal{V}_{-\frac{1}{2}}\left(z\right) & =\frac{1}{2}\mi k^{i}\me^{\frac{1}{2}\phi}\me^{\mi k\cdot
X}S_{\downarrow}^{-}\me^{\mi q\cdot H}\Sigma\left(z\right)\label{eq:pcos_on_vf_ext_sminus}\end{align} where\begin{eqnarray}
S_{\uparrow}^{-}=\me^{\mi\left(\frac{1}{2}H_{i}-\frac{1}{2}H_{j}\right)} & \qquad\text{and}\qquad &
S_{\downarrow}^{-}=\me^{\mi\left(-\frac{3}{2}H_{i}-\frac{1}{2}H_{j}\right)}\,.\label{eq:sminus_modified}\end{eqnarray} For the boson,
\begin{eqnarray} \lim_{w\rightarrow z}\me^{\phi}\partial\overline{X}^{i}\psi^{i}\left(w\right)\mathcal{V}_{-1}\left(z\right) & = &
\frac{1}{2}\mi\overline{k}^{i}\psi^{i}\me^{\mi k\cdot X}\me^{\mi q\cdot H}\Sigma\left(z\right)\nonumber \\ \lim_{w\rightarrow
z}\me^{\phi}\partial X^{i}\overline{\psi}^{i}\left(w\right)\mathcal{V}_{-1}\left(z\right) & = & \frac{1}{2}\mi
k^{i}\overline{\psi}^{i}\me^{\mi k\cdot X}\me^{\mi q\cdot H}\Sigma\left(z\right)\,.\label{eq:pcos_on_vb_ext}\end{eqnarray}

Since we should consider both possibilities for the boson, there are four possible terms from picture-changing the external dimensions:
two from the $H$-charge conserving combinations of (\ref{eq:pcos_on_vf_ext_splus}) and (\ref{eq:pcos_on_vb_ext}), and two from the
$H$-charge conserving combinations of (\ref{eq:pcos_on_vf_ext_sminus}) and (\ref{eq:pcos_on_vb_ext}).  The terms naturally group into two
pairs:\begin{multline} A_{2}^{+}=\frac{1}{4}\overline{u}_{1}u_{2}\mathcal{I}\left\langle
\me^{-\frac{1}{2}\phi}\left(z_{1}\right)\me^{\frac{1}{2}\phi}\left(z_{2}\right)\right\rangle \sum_{i\ne
j}^{2}\mi\overline{k}_{3}^{i}\left\langle \me^{\mi k_{1}\cdot X}\left(z_{1}\right)\partial X^{i}\me^{\mi k_{2}\cdot
X}\left(z_{2}\right)\me^{\mi k_{3}\cdot X}\left(z_{3}\right)\right\rangle \\ \left\langle \me^{-\frac{1}{2}\mi
H_{i}}\left(z_{1}\right)\me^{-\frac{1}{2}\mi H_{i}}\left(z_{2}\right)\me^{\mi H_{i}}\left(z_{3}\right)\right\rangle \left\langle
\me^{-\frac{1}{2}\mi H_{j}}\left(z_{1}\right)\me^{\frac{1}{2}\mi H_{j}}\left(z_{2}\right)\right\rangle \label{eq:a2plus}\end{multline}
\begin{multline}
A_{2}^{-}=\frac{1}{4}\overline{u}_{1}u_{2}\mathcal{I}\left\langle
\me^{-\frac{1}{2}\phi}\left(z_{1}\right)\me^{\frac{1}{2}\phi}\left(z_{2}\right)\right\rangle \sum_{i\ne j}^{2}\mi k_{3}^{i}\left\langle
\me^{\mi k_{1}\cdot X}\left(z_{1}\right)\partial X^{i}\me^{\mi k_{2}\cdot X}\left(z_{2}\right)\me^{\mi k_{3}\cdot
X}\left(z_{3}\right)\right\rangle \\ \left\langle \me^{\frac{1}{2}\mi H_{i}}\left(z_{1}\right)\me^{\frac{1}{2}\mi
H_{i}}\left(z_{2}\right)\me^{-\mi H_{i}}\left(z_{3}\right)\right\rangle \left\langle \me^{\frac{1}{2}\mi
H_{j}}\left(z_{1}\right)\me^{-\frac{1}{2}\mi H_{j}}\left(z_{2}\right)\right\rangle \,,\label{eq:a2minus}\end{multline} and\begin{multline}
A_{3}^{+}=-\frac{1}{4}\overline{u}_{1}u_{2}\mathcal{I}\left\langle
\me^{-\frac{1}{2}\phi}\left(z_{1}\right)\me^{\frac{1}{2}\phi}\left(z_{2}\right)\right\rangle \left\langle \me^{\mi k_{1}\cdot
X}\left(z_{1}\right)\me^{\mi k_{2}\cdot X}\left(z_{2}\right)\me^{\mi k_{3}\cdot X}\left(z_{3}\right)\right\rangle \\ \sum_{i\ne
j}^{2}k_{2}^{i}\overline{k}_{3}^{i}\left\langle \me^{-\frac{1}{2}\mi H_{i}}\left(z_{1}\right)\me^{\frac{3}{2}\mi
H_{i}}\left(z_{2}\right)\me^{-\mi H_{i}}\left(z_{3}\right)\right\rangle \left\langle \me^{-\frac{1}{2}\mi
H_{j}}\left(z_{1}\right)\me^{\frac{1}{2}\mi H_{j}}\left(z_{2}\right)\right\rangle \label{eq:a3plus}\end{multline}
\begin{multline}
A_{3}^{-}=-\frac{1}{4}\overline{u}_{1}u_{2}\mathcal{I}\left\langle
\me^{-\frac{1}{2}\phi}\left(z_{1}\right)\me^{\frac{1}{2}\phi}\left(z_{2}\right)\right\rangle \left\langle \me^{\mi k_{1}\cdot
X}\left(z_{1}\right)\me^{\mi k_{2}\cdot X}\left(z_{2}\right)\me^{\mi k_{3}\cdot X}\left(z_{3}\right)\right\rangle \\ \sum_{i\ne
j}^{2}\overline{k}_{2}^{i}k_{3}^{i}\left\langle \me^{\frac{1}{2}\mi H_{i}}\left(z_{1}\right)\me^{-\frac{3}{2}\mi
H_{i}}\left(z_{2}\right)\me^{\mi H_{i}}\left(z_{3}\right)\right\rangle \left\langle \me^{\frac{1}{2}\mi
H_{j}}\left(z_{1}\right)\me^{-\frac{1}{2}\mi H_{j}}\left(z_{2}\right)\right\rangle \,,\label{eq:a3minus}\end{multline} where the
contribution from the internal dimensions is\begin{equation} \mathcal{I}=\left\langle \me^{\mi q_{1}\cdot H}\left(z_{1}\right)\me^{\mi
q_{2}\cdot H}\left(z_{2}\right)\me^{\mi q_{3}\cdot H}\left(z_{3}\right)\right\rangle \left\langle
\Sigma_{1}\left(z_{1}\right)\Sigma_{2}\left(z_{2}\right)\Sigma_{3}\left(z_{3}\right)\right\rangle \,.\label{eq:a23_internal}\end{equation}

So, overall, we have five different terms contributing to $A$. Heuristically, the difference between them is that $A_{1}$ contains excited
twist operators but contains no kinematic factors, whereas the $A_{2}^{\pm}$ and $A_{3}^{\pm}$ terms contain only ordinary twist operators
but have kinematic factors in front of them. These kinematic factors will generally cause the $A_{2}^{\pm}$ and $A_{3}^{\pm}$ terms to be
suppressed relative to the $A_{1}$ contribution, unless we bring two of the vertex operators close together. Therefore, it seems
appropriate to identify the term $A_{1}$ with limit 1 in figure \ref{fig:limits}, and $A_{2}^{\pm},\, A_{3}^{\pm}$ with the other cases
\cite{Mandal:1987kz}.  When we compute these terms explicitly, we will see that this identification is indeed correct; first, however, we
must find explicit expressions for the correlation functions in $A_{1}$, $A_{2}$ and $A_{3}$.  The discussion is somewhat technical, so
for reference we have collected the results at the beginning of the section \ref{sec:quantum}, where we start to build the amplitudes.

\subsection{Bosonic fields\label{sub:corr_bosonic}}

Defining the annulus as the rectangular region of the complex plane $x\in\left[0,\frac{1}{2}\right]$, $y\in\left[0,t\right]$, the
correlators involving the fields $X^{\mu}$ may be found from those on the torus via the method of images \cite{Abel:2003ue}. The result
is\begin{equation} \left\langle X\left(z_{1}\right)\, X\left(z_{2}\right)\right\rangle
=-\frac{1}{2}\left|\log\vartheta_{1}\left(z_{2}-z_{1}|\mi
t\right)\right|-\frac{1}{2}\left|\log\vartheta_{1}\left(z_{2}+\overline{z}_{1}|\mi
t\right)\right|+r\left(t\right)\,,\label{eq:xx_torus}\end{equation} with the Jacobi theta function $\vartheta_{1}$ defined as in appendix
\ref{sec:theta}, and $r\left(t\right)$ a suitable function which regulates the propagator as $z_{1}\rightarrow z_{2}$. All of our fields
are on the same end of the annulus, so we can set $z_{1}=\mi y_{1}$, $z_{2}=\mi y_{2}$ with $y_{1},\, y_{2}\in[0,t]$. Then, since the
theta functions are real for purely imaginary arguments,\begin{equation} \left\langle X\left(y_{1}\right)\,
X\left(y_{2}\right)\right\rangle =-\log\vartheta_{1}\left(\mi y_{12}\right)\,,\label{eq:xx_annulus}\end{equation} with the notation
$y_{ij}=\left|y_{j}-y_{i}\right|$. For brevity, we have suppressed the explicit $t$-dependence in $\vartheta_{1}$.  We have also left out
the function $r\left(t\right)$; it will not become important until we begin to care about the $t$-dependence of $A$ in section
\ref{sec:beta}, and we will deal with it then.

With this simplification, the $\me^{\mi k\cdot X}$ correlations in $A$ are\begin{eqnarray} \left\langle \prod_{i}\me^{\mi k_{i}\cdot
X}\left(y_{i}\right)\right\rangle & = & \prod_{i<j}\vartheta_{1}\left(\mi y_{ij}\right)^{k_{i}\cdot
k_{j}}\,,\label{eq:eikx_corr}\end{eqnarray} while including one factor of $\partial X^{\mu}$ pulls down kinematic factors:\begin{equation}
\left\langle \partial X^{\mu}\left(y_{k}\right)\prod_{i}\me^{\mi k_{i}\cdot X}\left(y_{i}\right)\right\rangle
=\prod_{\stackrel{{\scriptstyle i<j}}{{\scriptstyle i\ne k}}}\mi k_{j}^{\mu}\frac{\theta_{1}^{\prime}\left(\mi
y_{jk}\right)}{\theta_{1}\left(\mi y_{jk}\right)}\vartheta_{1}\left(\mi y_{ij}\right)^{k_{i}\cdot
k_{j}}\,.\label{eq:dxeikx_corr}\end{equation}

\subsection{Spin fields\label{sub:corr_spin}}

The correlators between the the fermion spin fields $\me^{\mi q\cdot H}$ are more difficult to compute. The problem is that we must
respect spin structures on the underlying torus, which prevents us from dealing with the $H$ fields in the same way as the $X$ fields. We
may resolve the issue by using the stress-tensor method~\cite{Dixon:1987qv}, generalising the results of ref.~\cite{Atick:1987ns}.

As above, we first perform the calculation on a torus, and then specialize the result to the annulus. Begin by bosonizing the holomorphic
fermions,\begin{equation} \psi\left(z\right)=\me^{\mi H\left(z\right)}\qquad\overline{\psi}\left(z\right)=\me^{-\mi
H\left(z\right)}\,,\label{eq:fermion_bos}\end{equation} and defining holomorphic spin operators,\begin{equation}
\mathcal{S}_{a}\left(z\right)=\me^{\mi aH\left(z\right)}\,.\label{eq:spin_ops}\end{equation} From (\ref{eq:opes_xpsi}), we see the the
OPEs between fermions and spin fields take the form\begin{equation}
\begin{split}\psi\left(z\right)\,\overline{\psi}\left(w\right) & \sim\left(z-w\right)^{-1}\\
\psi(z)\,\mathcal{S}_{a}(w) & \sim\left(z-w\right)^{a}\,\mathcal{S}_{a+1}\left(w\right)\\
\overline{\psi}(z)\,\mathcal{S}_{a}(w) & \sim\left(z-w\right)^{-a}\,\mathcal{S}_{a-1}\left(w\right)\\
\mathcal{S}_{a}(z)\,\mathcal{S}_{b}(w) & \sim\left(z-w\right)^{ab}\,\mathcal{S}_{a+b}\left(w\right)\,.\end{split}
\label{eq:spin_opes}\end{equation}
We will work out the correlation between an arbitrary number of spin fields, since it is no more complicated than that between three such
fields. Define an auxiliary Green's function as\begin{equation} g\left(z,w;z_{i}\right)=\frac{\left\langle
\overline{\psi}\left(z\right)\psi\left(w\right)\prod_{i}S_{a_{i}}\left(z_{i}\right)\right\rangle }{\left\langle
\prod_{i}S_{a_{i}}\left(z_{i}\right)\right\rangle }\,.\label{eq:g_spin_defn}\end{equation} The OPEs (\ref{eq:spin_opes}) show that this
function must satisfy\begin{align} g\left(z,w;z_{i}\right) & =\left(z-w\right)^{-1}+\text{finite as $z\rightarrow w$}\nonumber \\
g\left(z,w;z_{i}\right) & \propto\left(z-w\right)^{-a_{i}}+\text{finite as $z\rightarrow z_{i}$}\label{eq:g_behaviour}\\
g\left(z,w;z_{i}\right) & \propto\left(z-w\right)^{a_{i}}+\text{finite as $w\rightarrow z_{i}$}.\nonumber \end{align} Furthermore, it must
be periodic on the torus as a function of $z$ and $w$. A suitable function satisfying these conditions is\begin{equation}
g\left(z,w;z_{i}\right)=\frac{\vartheta_{1}'\left(0\right)}{\vartheta_{1}\left(z-w\right)}\frac{\vartheta_{\alpha\beta}\left(z-w+\sum
a_{i}z_{i}\right)}{\vartheta_{\alpha\beta}\left(\sum
a_{i}z_{i}\right)}\prod_{i}\left(\frac{\vartheta_{1}\left(w-z_{i}\right)}{\vartheta_{1}\left(z-z_{i}\right)}\right)^{a_{i}}\,,\label{eq:spin_g}\end{equation}
where $\left(\alpha\beta\right)=\left(00\right)\ldots\left(11\right)$ label the four possible spin structures on the torus. The reason for
the specific form of the term $\sum_{i=1}^{n}a_{i}z_{i}$ in the argument of $\vartheta_{\alpha\beta}$ will become clear shortly; for the
moment we press on.

The stress-energy tensor for the $\psi,\,\overline{\psi}$ CFT may be written as\begin{equation} T(z)=\lim_{z\rightarrow
w}\left[\frac{1}{2}\partial_{z}\overline{\psi}\left(z\right)\psi\left(w\right)-\frac{1}{2}\overline{\psi}\left(z\right)\partial_{w}\psi\left(w\right)+\frac{1}{\left(z-w\right)^{2}}\right]\,,\label{eq:sem_tensor}\end{equation}
and so using the definition (\ref{eq:g_spin_defn}), we may form the function\begin{equation} \frac{\left\langle
T\left(z\right)\prod_{i}S_{a_{i}}\left(z_{i}\right)\right\rangle }{\left\langle \prod_{i}S_{a_{i}}\left(z_{i}\right)\right\rangle
}=\lim_{z\rightarrow
w}\left[\frac{1}{2}\partial_{z}g\left(z,w;z_{i}\right)-\frac{1}{2}\partial_{w}g\left(z,w;z_{i}\right)+\frac{1}{\left(z-w\right)^{2}}\right]\,.\label{eq:spin_stress}\end{equation}
Using $\lim_{z\rightarrow0}\vartheta_{1}\left(z\right)=\vartheta_{1}^{\prime}\left(0\right)z$, and derivatives thereof, one finds\[
\frac{\left\langle T\left(z\right)\prod_{i}S_{a_{i}}\left(z_{i}\right)\right\rangle }{\left\langle
\prod_{i}S_{a_{i}}\left(z_{i}\right)\right\rangle
}=\frac{1}{2}\left[\sum_{i=1}^{n}a_{i}\frac{\vartheta_{1}^{\prime}\left(z-z_{i}\right)}{\vartheta_{1}\left(z-z_{i}\right)}\right]^{2}-\left[\sum_{i=1}^{n}a_{i}\frac{\vartheta_{1}^{\prime}\left(z-z_{i}\right)}{\vartheta_{1}\left(z-z_{i}\right)}\right]\frac{\vartheta_{\alpha\beta}^{\prime}\left(\sum_{i}a_{i}z_{i}\right)}{\vartheta_{\alpha\beta}\left(\sum_{i}a_{i}z_{i}\right)}\,.\]

The next step is to take the limit $z\rightarrow z_{j}$ on both sides.  In this limit, the right-hand side may be evaluated directly,
whilst the left-hand side is simplified by the fact that the OPE of any operator with the stress-energy tensor takes the prescribed
form\begin{equation}
T\left(z\right)\,\mathcal{S}_{a_{j}}\left(z_{j}\right)\sim\frac{h}{\left(z-z_{j}\right)^{2}}\mathcal{S}_{a_{j}}\left(z_{j}\right)+\frac{1}{z-z_{j}}\partial_{z_{j}}\mathcal{S}_{a_{j}}\left(z_{j}\right)\,,\end{equation}
with $h$ the conformal weight of the field $\mathcal{S}_{a_{j}}\left(z\right)$.  The result is\begin{multline} \lim_{z\rightarrow
z_{j}}\left\{ \frac{h}{\left(z-z_{j}\right)^{2}}+\frac{1}{z-z_{j}}\partial_{z_{j}}\log\left\langle
\prod_{i}\mathcal{S}_{a_{i}}\left(z_{i}\right)\right\rangle \right\} \\ =\lim_{z\rightarrow z_{j}}\left\{
\frac{\frac{1}{2}a_{j}^{2}}{(z-z_{j})^{2}}+\frac{a_{j}}{z-z_{j}}\left[\left(\sum_{i\ne
j}a_{i}\frac{\vartheta_{1}'(z-z_{i})}{\vartheta_{1}(z-z_{i})}\right)-\frac{\vartheta_{\alpha\beta}'\left(\sum_{i}a_{i}z_{i}\right)}{\vartheta_{\alpha\beta}\left(\sum_{i}a_{i}z_{i}\right)}\right]\right\}
\,.\label{eq:spin_lim}\end{multline} Comparing coefficients in $(z-z_{j})$ shows that $\mathcal{S}_{a_{j}}(z)$ correctly has conformal
weight $\frac{1}{2}a_{j}^{2}$, and that\begin{equation} \partial_{z_{j}}\log\left\langle
\prod_{i}\mathcal{S}_{a_{i}}\left(z_{i}\right)\right\rangle =a_{j}\left[\left(\sum_{i\ne
j}a_{i}\frac{\vartheta_{1}'\left(z-z_{i}\right)}{\vartheta_{1}\left(z-z_{i}\right)}\right)-\frac{\vartheta_{\alpha\beta}'\left(\sum_{i}a_{i}z_{i}\right)}{\vartheta_{\alpha\beta}\left(\sum_{i}a_{i}z_{i}\right)}\right]\,,\label{eq:spin_diff}\end{equation}
with solution\begin{equation} \left\langle \prod_{i}\mathcal{S}_{a_{i}}\left(z_{i}\right)\right\rangle
=K_{\alpha\beta}\left[\prod_{i<j}\vartheta_{1}\left(z_{i}-z_{j}\right)^{a_{i}a_{j}}\right]\vartheta_{\alpha\beta}\left(\sum_{i}a_{i}z_{i}\right)\label{eq:spin_corr}\end{equation}
where $K_{\alpha\beta}$ is an overall normalization.

The reason for the argument of $\vartheta_{\alpha\beta}$ in (\ref{eq:spin_g}) to take the form it does is now revealed: when
$a_{i}=\pm\frac{1}{2}$, the three translations $z_{i}\rightarrow z_{i}+1$, $z_{i}\rightarrow z_{i}+\tau$ and $z_{i}\rightarrow
z_{i}+1+\tau$ transform a given theta function $\vartheta_{\alpha\beta}$ into the other three theta functions. Under the same
translations, the correlator (\ref{eq:spin_corr}) for a given spin structure transforms into the correlator for one of the other spin
structures (up to a phase), just as it should. For the internal spin fields, on the other hand, translational invariance implies
$\sum_{i}a_{i}=0$, which is just the $H$-charge conservation condition that we used in section \ref{sub:vertex_and_pcos}.

We can convert our result for general spin operators on the torus to that for the annulus quite easily, by defining an open-string
auxiliary Green's function through the method of images,\begin{equation}
g_{\text{open}}\left(z,w;z_{i}\right)=g\left(z,w;z_{i}\right)+g\left(z,\bar{w};z_{i}\right)\,,\end{equation} and proceeding through with
the calculation as before. The outcome is that we end up with two identical copies of equation \ref{eq:spin_diff}; we see, then, that the
correlator for spin fields on the annulus is just the same as that on the torus.

\subsection{Ghost spin fields\label{sub:corr_ghost}}

Also required are correlators for the ghost spin fields, $\me^{a\phi}$.  These may be calculated by an analogous method, but with the OPEs
and stress tensor appropriately modified to account for the fact that the ghost fields $\phi$ inhabit the $\left(\beta,\,\gamma\right)$
rather than the $\left(\overline{\psi},\,\psi\right)$ CFT \cite{Atick:1986rs}.  The result turns out to be the reciprocal of
(\ref{eq:spin_corr}):\begin{equation} \left\langle \me^{a_{i}\phi}\left(z_{i}\right)\right\rangle
=K_{\alpha\beta}^{g}\left[\prod_{i<j}\vartheta_{1}\left(z_{i}-z_{j}\right)^{-a_{i}a_{j}}\right]\vartheta_{\alpha\beta}^{-1}\left(\sum_{i}a_{i}z_{i}\right)\,.\label{eq:ghost_corr}\end{equation}

\subsection{Twist fields\label{sub:corr_twist}}

Twist fields correlators are also tricky, this time because they do not possess a natural interpretation as local operators on the
worldsheet.  Again, the resolution is to use the stress-tensor method, and again we begin on a torus before converting our result to the
annulus.

On a torus, the twist fields have the OPEs (\ref{eq:opes_xsigma}), given by the local monodromy conditions for the twist fields. Things
are complicated somewhat by the \emph{global} monodromy conditions -- that is, the behaviour of the $X$ fields as they are transported
around collections of twist fields. If we define a closed loop $\gamma$ as a loop on the worldsheet as a loop enclosing twist operators
whose twists sum to zero, then around such a loop,\begin{equation} \Delta_{\gamma}X=\oint_{\gamma}\sdif{z}\partial
X+\oint_{\gamma}\sdif{\bar{z}}\overline{\partial}X=v_{\gamma}\,.\label{eq:twist_loop}\end{equation} Here $v_{\gamma}$ represents a
consistent displacement on the network of branes. If the branes are on a compact space then the contours may generate a displacement that
wraps a number of times around the compactified brane before returning to the same place. We may choose to split $X$ into a `classical'
part and a `quantum fluctuation' part,\begin{equation} X=X_{\text{cl}}+X_{\text{qu}}\,,\label{eq:twist_x_clqu}\end{equation} with the
requirement that the quantum part be unchanged by transportation around a closed loop while the classical part takes care of the
displacement,\begin{eqnarray} \Delta_{\gamma}X_{\text{qu}} & = & \oint_{\gamma}\sdif{z}\partial
X_{\text{qu}}+\oint_{\gamma}\sdif{\bar{z}}\overline{\partial}X_{\text{qu}}=0\,.\label{eq:twist_qu}\\ \Delta_{\gamma}X_{\text{cl}} & = &
\oint_{\gamma}\sdif{z}\partial X_{\text{cl}}+\oint_{\gamma}\sdif{\bar{z}}\overline{\partial}X_{\text{cl}}=v_{\gamma}\,.\end{eqnarray}
Then, the correlator for $L$ twist operators also splits into a quantum and a classical part,\begin{equation} \left\langle
\prod_{i=1}^{L}\sigma_{i}\left(z_{i}\right)\right\rangle
=Z_{\sigma}\me^{-S_{\text{cl}}}\,.\label{eq:twist_quantum_classical}\end{equation} The classical portion of the twist correlators will
play a key role in our calculation, and as such we defer discussion of it until section \ref{sec:classical}, considering first just the
quantum part $Z_{\sigma}$.

Using the stress-tensor method to compute this correlator in a manner consistent with both the local and global monodromy conditions, plus
the periodicity of the torus, is not a trivial task. Fortunately such a calculation has been performed on the torus by Atick \emph{et
al.}~\cite{Atick:1988kd}, with the result
\footnote{The OPEs in \cite{Atick:1988kd} have $\theta\rightarrow1-\theta$ with respect to our OPEs (\ref{eq:opes_xsigma}); in the
conventional orbifold language, we are interested in the correlator of anti-twist operators, rather than twist operators. We deal with the
disparity by reversing the notions of holomorphic and anti-holomorphic with respect to the work of Atick \emph{et al.}, leading to the
result shown in eq.~\ref{eq:twist_atick}. Formally, we should also replace $W\rightarrow\overline{W}$ everywhere; however, this does not
affect our results, so we stick with the simpler notation.}\begin{multline} Z_{\sigma}=\left|\det
W\right|^{-1}\overline{\vartheta_{1}\left(Y\right)}^{(L-M-1)}\vartheta_{1}\left(Y^{\prime}\right)^{(M-1)}\prod_{i<j}^{L-M}\overline{\vartheta_{1}\left(z_{\alpha_{i}}-z_{\alpha_{j}}\right)}\prod_{i<j}^{M}\vartheta_{1}\left(z_{\beta_{i}}-z_{\beta_{j}}\right)\\
\prod_{i<j}^{L}\overline{\vartheta_{1}\left(z_{i}-z_{j}\right)}^{-(1-\theta_{i})(1-\theta_{j})}\vartheta_{1}\left(z_{i}-z_{j}\right)^{-\theta_{i}\theta_{j}}\,,\label{eq:twist_atick}\end{multline}
where $M=\sum_{i=1}^{L}\theta_{i}$ is an integer, $z_{\alpha_{i}}$ are a set of $L-M$ twist insertion points chosen from $z_{i}$ and
$z_{\beta_{j}}$ are a set of $M$ twist insertion points chosen from $z_{i}$ (not necessarily related to $z_{\alpha_{i}}$). The elements of
the elements of the $L\times L$ matrix $W$ are given by\begin{eqnarray} W_{a}^{i} & = &
\oint_{\gamma_{a}}\sdif{z}\omega_{i}\left(z\right)\qquad i=1,\ldots,L-M\nonumber \\ W_{a}^{i} & = &
\oint_{\gamma_{a}}\sdif{\bar{z}}\overline{\omega_{i}^{\prime}\left(\overline{z}\right)},\qquad
i=L-M+1,\ldots,L\,,\label{eq:twist_wmatrix}\end{eqnarray} with $\gamma_{a}$ a basis for $L$ closed loops on the worldsheet.  $\omega$ and
$\omega^{\prime}$ are so-called `cut differentials',\begin{align} \omega_{i}\left(z\right)= &
\gamma\left(z\right)\vartheta_{1}\left(z-z_{\alpha_{i}}-Y\right)\prod_{i\ne j}^{L-M}\vartheta_{1}\left(z-z_{\alpha_{j}}\right)\nonumber \\
\omega_{i}^{\prime}\left(z\right)= & \gamma^{\prime}\left(z\right)\vartheta_{1}\left(z-z_{\beta_{i}}-Y'\right)\prod_{i\ne
j}^{M}\vartheta_{1}\left(z-z_{\beta_{j}}\right)\label{eq:twist_cutdiff}\end{align} with the contributions\begin{align} \gamma(z)= &
\prod_{j=1}^{L}\vartheta_{1}\left(z-z_{j}\right)^{-\left(1-\theta_{j}\right)}\nonumber \\ \gamma^{\prime}(z)= &
\prod_{j=1}^{L}\vartheta_{1}\left(z-z_{j}\right)^{-\theta_{j}}\label{eq:twist_gamma}\end{align} chosen to obey the local monodromy. These
cut differentials form a basis for $\partial X_{\text{cl}}$;\begin{eqnarray} \partial\overline{X}_{\text{cl}} & = &
\sum_{i=1}^{L-M}c_{i}\omega_{i}\nonumber \\ \partial X_{\text{cl}} & = &
\sum_{i=L-M+1}^{L}c_{i}\omega_{i}^{\prime}\,.\label{eq:twist_dX}\end{eqnarray} The terms $Y$ and $Y^{\prime}$ in (\ref{eq:twist_cutdiff})
are\begin{eqnarray} Y & = & \sum_{i=1}^{L}(1-\theta_{i})z_{i}-\sum_{j=1}^{L-M}z_{\alpha_{j}}\nonumber \\ Y^{\prime} & = &
\sum_{i=1}^{L}\theta_{i}z_{i}-\sum_{j=1}^{M}z_{\beta_{j}}\,,\label{eq:twist_y}\end{eqnarray} and have the function of keeping $\omega$ and
$\omega^{\prime}$ periodic on the torus.

This result may be converted to the annulus by applying the method of images, just as for the spin fields. This time, the two terms in the
open-string Green's function lead to \emph{different} differential equations; this is a reflection of the fact that the twist operators
respond differently to holomorphic and anti-holomorphic fields on the torus, as we see from the OPEs above. Taking the vertex operators to
lie along the annulus boundary, $z_{i}=\mi y_{i}$, we find that both differential equations are in the same variable and so we may add
them before solving. The result turns out to be the square root of (\ref{eq:twist_atick}), which is exactly what similar calculations at
tree level~\cite{Abel:2003vv,Abel:2003yx,Cvetic:2003ch} lead us to expect.

We are interested in the case where we have three twist operators, and the angles at the intersections add up to $\pi$; hence, $L=3$ and
$M=1$. We choose the $3-1=2$ points $z_{\alpha_{i}}$ as $\{ z_{\alpha_{1}},z_{\alpha_{2}}\}=\{\mi y_{1},\mi y_{2}\}$, and the one point
$z_{\beta_{1}}=\mi y_{3}$. Then, for a single complex dimension and up to an overall normalization,
\begin{multline}
Z_{\sigma}=\left|\det W\right|^{-\frac{1}{2}}\vartheta_{1}\left(\mi\left(\theta_{1}y_{1}+\theta_{2}y_{2}-(1-\theta_{3})y_{3}\right)\right)^{\frac{1}{2}}\vartheta_{1}\left(\mi y_{12}\right)^{\frac{1}{2}}\\
\prod_{i<j}^{3}\vartheta_{1}\left(\mi y_{ij}\right)^{-\frac{1}{2}\left(1-\theta_{i}\right)\left(1-\theta_{j}\right)-\frac{1}{2}\theta_{i}\theta_{j}}\label{eq:twist_zqu}
\end{multline}
(an overall phase having been taken out of the first $\vartheta_{1}$), with\begin{align} W_{\alpha}^{1}= &
\oint_{\gamma_{\alpha}}\sdif{z}\gamma\left(z\right)\vartheta_{1}\left(z-\left(1-\theta_{1}\right)\mi y_{1}+\theta_{2}\mi
y_{2}-\left(1-\theta_{3}\right)y_{3}\right)\vartheta_{1}\left(z-\mi y_{2}\right)\nonumber \\ W_{\alpha}^{2}= &
\oint_{\gamma_{\alpha}}\sdif{z}\gamma\left(z\right)\vartheta_{1}\left(z+\theta_{1}\mi
y_{1}-\left(1-\theta_{2}\right)y_{2}-\left(1-\theta_{3}\right)y_{3}\right)\vartheta_{1}\left(z-\mi y_{1}\right)\nonumber \\
W_{\alpha}^{3}= &
\oint_{\gamma_{\alpha}}\sdif{\bar{z}}\overline{\gamma^{\prime}\left(\bar{z}\right)}\overline{\vartheta_{1}\left(\bar{z}-\mi\left(\theta_{1}y_{1}+\theta_{2}y_{2}+\theta_{3}y_{3}\right)\right)}\,.\label{eq:twist_wigrls}\end{align}

\FIGURE[t]{\includegraphics{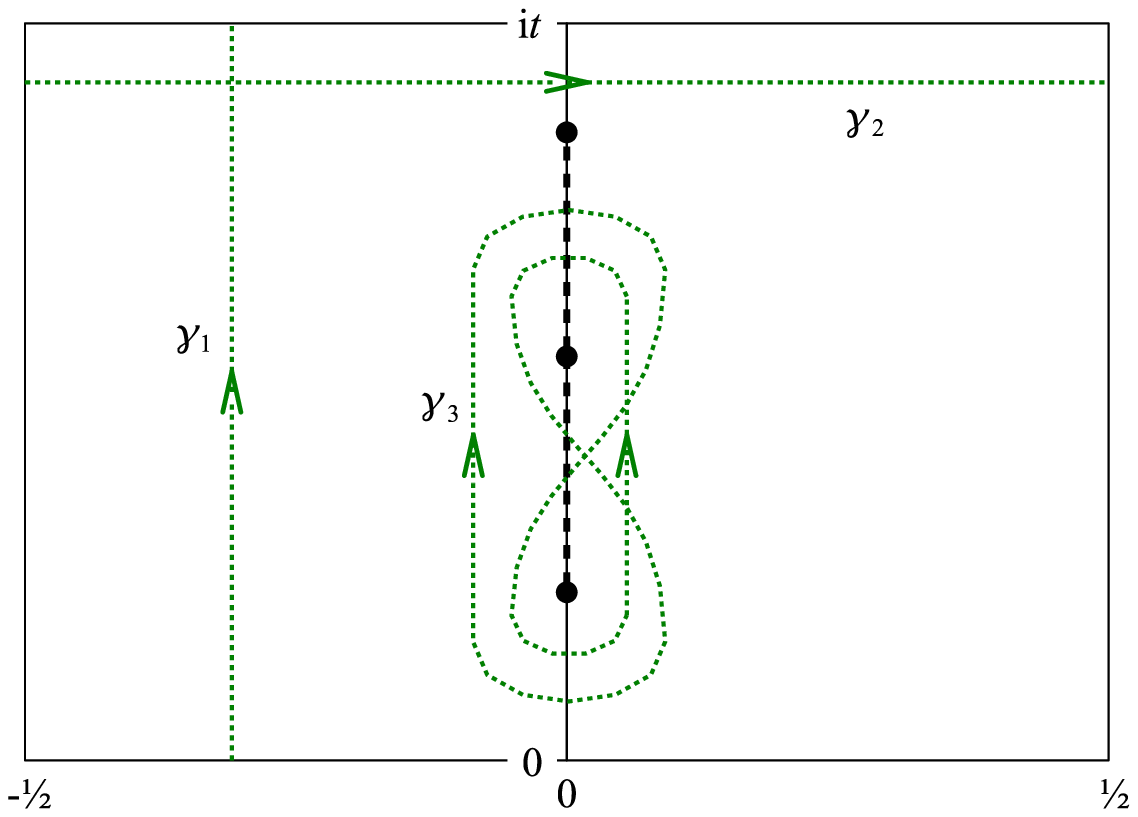}
\caption{Independent contours on the worldsheet: two cycles and a Pochhammer loop. The annulus corresponds to the region
$z>0$.\label{fig:contours}} } Three suitable independent contours $\gamma_{a}$ are shown in figure \ref{fig:contours}. They consist of a
cycle $\gamma_{1}$ running around the annulus, a cycle $\gamma_{2}$ around the covering torus and a Pochhammer contour $\gamma_{3}$. This
latter contour encircles each of the twist operators once in each direction, so that the net twist enclosed by $\gamma_{3}$ is correctly
zero. We choose $\gamma_{3}$ to encircle the branch cut $\mi y_{1}<z<\mi y_{2}$, but we could just have easily have chosen it to encircle
the other cut; indeed, one may deform between the two by taking linear combinations of $\gamma_{3}$ together with the two cycles.

Note that the cycle $\gamma_{2}$, which generates a displacement between the branes, is always chosen such that it does not pass through a
branch cut. This corresponds to taking the displacement to be between two parallel pairs of branes.

\subsection{Excited twist fields\label{sub:corr_extwist}}

Finally, we must deal with the excited twist correlator $\left\langle
\sigma\left(z_{1}\right)\tau\left(z_{2}\right)\tau^{\prime}\left(z_{3}\right)\right\rangle $ which is present in $A_{1}$
(eq. \ref{eq:a1}). This correlator may be obtained from the basic twist correlator as follows: using the OPEs
(\ref{eq:opes_xsigma}), write\begin{equation} \left\langle
\sigma\left(z_{1}\right)\tau\left(z_{2}\right)\tau^{\prime}\left(z_{3}\right)\right\rangle =\lim_{\stackrel{{\scriptstyle
z\rightarrow z_{2}}}{w\rightarrow z_{3}}}\left(z-z_{2}\right)^{\theta_{2}}\left(w-z_{3}\right)^{1-\theta_{3}} \left\langle
\sigma\left(z_{1}\right)\partial
X\left(z\right)\sigma\left(z_{2}\right)\partial\overline{X}\left(w\right)\sigma\left(z_{3}\right)\right\rangle
\,.\label{eq:extwist_opes}
\end{equation}
Writing $X=X_{\text{qu}}+X_{\text{cl}}$, this second correlator splits into a quantum and a
classical piece,\begin{multline} \left\langle \sigma\left(z_{1}\right)\partial
X\left(z\right)\sigma\left(z_{2}\right)\partial\overline{X}\left(w\right)\sigma\left(z_{3}\right)\right\rangle =\left\langle
\sigma\left(z_{1}\right)\partial
X_{\text{qu}}\left(z\right)\sigma\left(z_{2}\right)\partial\overline{X}_{\text{qu}}\left(w\right)\sigma\left(z_{3}\right)\right\rangle \\
+\partial X_{\text{cl}}\left(z\right)\partial\overline{X}_{\text{cl}}\left(w\right)\left\langle
\sigma\left(z_{1}\right)\sigma\left(z_{2}\right)\sigma\left(z_{3}\right)\right\rangle
\,,\label{eq:extwist_quantum_classical}\end{multline} and we deal with each portion separately.

Taking the quantum part first, define a function $g\left(z,w;z_{i}\right)$ as\begin{equation} g\left(z,w;z_{i}\right)=\frac{\left\langle
\partial X_{\text{qu}}\left(z\right)\partial\overline{X}_{\text{qu}}\left(w\right)\prod_{i=1}^{3}\sigma\left(z_{i}\right)\right\rangle
}{\left\langle \prod_{i=1}^{3}\sigma\left(z_{i}\right)\right\rangle }\,.\label{eq:extwist_g}\end{equation} so that\begin{equation}
\left\langle \sigma_{1}\left(z_{1}\right)\tau_{2}\left(z_{2}\right)\tau_{3}^{\prime}\left(z_{3}\right)\right\rangle
_{\text{qu}}=\left\langle \prod_{i}\sigma\left(z_{i}\right)\right\rangle \lim_{\stackrel{{\scriptstyle z\rightarrow z_{2}}}{w\rightarrow
z_{3}}}\left(z-z_{2}\right)^{\theta_{2}}\left(w-z_{3}\right)^{1-\theta_{3}}g\left(z,w;z_{i}\right)\,.\label{eq:extwist_limg}\end{equation}
Since $g$ is a function only of holomorphic variables, this result will be equally true on the torus or annulus; there is no need to take
square roots anywhere except in the base twist correlator $\left\langle \prod_{i}\sigma_{i}\left(z_{i}\right)\right\rangle $.

An explicit expression for $g\left(z,w;z_{i}\right)$ is\footnote{As in the previous section, we deal with the $\theta\rightarrow1-\theta$
issue by exchanging $X\leftrightarrow\overline{X}$, which corresponds simply to swapping $z\leftrightarrow w$ relative to the work of
Atick \emph{et al.}} \cite{Atick:1988kd}\begin{align} g\left(z,w;z_{i}\right)= &
g_{s}\left(z,w\right)+\omega_{1}\left(w\right)\sum_{i=1}^{2}B_{i}\omega_{i}^{\prime}\left(z\right)\,.\label{eq:extwist_g2}\end{align}
Using the result $\lim_{z\rightarrow0}\vartheta_{1}\left(z\right)=\vartheta_{1}^{\prime}\left(0\right)z$, we see\begin{align}
\lim_{w\rightarrow z_{3}}\left(w-z_{3}\right)^{1-\theta_{3}}\omega_{i}\left(w\right) & \propto\lim_{w\rightarrow
z_{3}}\left(w-z_{3}\right)^{1-\theta_{3}}\vartheta_{1}\left(w-z_{3}\right)^{-\left(1-\theta_{3}\right)}\vartheta_{1}\left(w-z_{3}\right)\label{eq:extwist_vanish}\\
& =0\,,\nonumber \end{align} and so the second term in $g\left(z,w;z_{i}\right)$ disappears when we take the limit in equation
\ref{eq:extwist_limg}. Therefore, the term of interest to us is $g_{s}\left(z,w\right)$, which is given by\begin{equation}
g_{s}\left(z,w\right)=\gamma\left(w\right)\gamma^{\prime}\left(z\right)\left[\frac{\vartheta_{1}^{\prime}\left(0\right)}{\vartheta_{1}\left(w-z\right)}\right]^{2}\sum_{i=1}^{3}\theta_{i}F_{i}\left(w,z\right)\vartheta_{1}\left(z-z_{i}\right)\prod_{j\ne
i}\vartheta_{1}\left(w-z_{j}\right)\,,\label{eq:extwist_gs}\end{equation} with\begin{align} F_{i}\left(w,z\right)= &
\frac{\vartheta_{1}\left(w-z+U_{i}^{0}\right)}{\vartheta_{1}\left(U_{i}^{0}\right)}\frac{\vartheta_{1}\left(w-z+Y_{i}-U_{i}^{0}\right)}{\vartheta_{1}\left(Y_{i}-U_{i}^{0}\right)}\label{eq:extwist_f}\\
Y_{i}= & \left(\sum_{j=1}^{3}\theta_{j}z_{j}\right)-z_{i}\,,\label{eq:extwist_y}\end{align} and $U_{i}^{0}$ chosen so as to satisfy the
equation $\partial_{w}F_{i}\left(w,z\right)=0$.  Taking the limit (\ref{eq:extwist_limg}), one finds
\begin{multline}
\left\langle \sigma_{1}\left(z_{1}\right)\tau_{2}\left(z_{2}\right)\tau_{3}^{\prime}\left(z_{3}\right)\right\rangle _{\text{qu}}=\left\langle \prod_{i}\sigma_{i}\left(z_{i}\right)\right\rangle \theta_{3}F_{3}\left(z_{3},z_{2}\right)\\
\vartheta_{1}\left(z_{12}\right)^{-\theta_{1}}\vartheta_{1}\left(z_{13}\right)^{\theta_{1}}\vartheta_{1}\left(z_{23}\right)^{\theta_{2}-\theta_{3}-1}\,.\label{eq:extwist_quantum}
\end{multline}

To evaluate the classical part of (\ref{eq:extwist_quantum_classical}) we insert (\ref{eq:twist_dX}) for the case of three twist
operators,\begin{eqnarray} \partial\overline{X}_{\text{cl}} & = & \sum_{i=1}^{2}c_{i}\omega_{i}\nonumber \\ \partial X_{\text{cl}} & = &
c_{3}\omega_{3}^{\prime}\,,\label{eq:extwist_class_dX}\end{eqnarray} and take the limit (\ref{eq:extwist_limg}) to find\begin{multline}
\left\langle \sigma_{1}\left(z_{1}\right)\tau_{2}\left(z_{2}\right)\tau_{3}^{\prime}\left(z_{3}\right)\right\rangle
_{\text{cl}}=\left\langle \prod_{i}\sigma_{i}\left(z_{i}\right)\right\rangle
\vartheta_{1}\left(z_{12}\right)^{-\theta_{1}}\vartheta_{1}\left(z_{13}\right)^{\theta_{1}-1}\vartheta_{1}\left(z_{23}\right)^{\theta_{2}-\theta_{3}-1}\\
\left[c_{1}\vartheta_{1}\left(z_{23}\right)\vartheta_{1}\left(-z_{1}\left(1-\theta_{1}\right)+z_{2}\theta_{2}+z_{3}\theta_{3}\right)+c_{2}\vartheta_{1}\left(z_{13}\right)\vartheta_{1}\left(z_{1}\theta_{1}+z_{2}\theta_{2}-z_{3}\left(1-\theta_{3}\right)\right)\right]\\
c_{3}\vartheta_{1}\left(z_{1}\theta_{1}-z_{2}\left(1-\theta_{2}\right)+z_{3}\theta_{3}\right)\,.\label{eq:extwist_classical}\end{multline}

The factors of $\vartheta_{1}\left(z_{ij}\right)^{\pm\theta}$ in both the quantum and classical parts of this excited twist correlator
compensate exactly for the changes in the spin correlator introduced by picture-changing.

\section{The quantum part\label{sec:quantum}}

We are now in a position to construct the quantum part of $A$ explicitly.  In everything that follows it should be understood that the
vertex operators $z_{i}$ are positioned along the imaginary axis as shown, so that $z_{i}=\mi y_{i}$. For ease of reference, we collect
together the relevant correlators (\ref{eq:eikx_corr}), (\ref{eq:dxeikx_corr}), (\ref{eq:spin_corr}), (\ref{eq:ghost_corr}),
(\ref{eq:twist_zqu}), (\ref{eq:extwist_quantum}) and (\ref{eq:extwist_classical}) in figure \ref{fig:correlators}.
\afterpage{\clearpage}
\FIGURE{\parbox{\hsize}{\begin{center}
\begin{tabular}{|p{155mm}|}
\hline 
~\\~\\\begin{minipage}[c]{1.0\columnwidth}
\begin{eqnarray*}
\left\langle \prod_{i}\me^{\mi k_{i}\cdot X}\left(z_{i}\right)\right\rangle  & = & \prod_{i<j}\vartheta_{1}\left(z_{ij}\right)^{k_{i}\cdot k_{j}}\\
\left\langle \partial X^{\mu}\left(w\right)\prod_{i}\me^{\mi k_{i}\cdot X}\left(z_{i}\right)\right\rangle  & = & \prod_{i<j}\mi k_{j}^{\mu}\frac{\theta_{1}^{\prime}\left(w-z_{j}\right)}{\theta_{1}\left(w-z_{j}\right)}\vartheta_{1}\left(z_{ij}\right)^{k_{i}\cdot k_{j}}\\
\left\langle \prod_{i}\me^{\mi a_{i}H}\left(z_{i}\right)\right\rangle  & = & K_{\alpha\beta}\left[\prod_{i<j}\vartheta_{1}\left(z_{i}-z_{j}\right)^{a_{i}a_{j}}\right]\vartheta_{\alpha\beta}\left(\sum_{i}a_{i}z_{i}\right)\\
\left\langle \prod_{i}\me^{a_{i}\phi}\left(z_{i}\right)\right\rangle  & = & K_{\alpha\beta}^{g}\left[\prod_{i<j}\vartheta_{1}\left(z_{i}-z_{j}\right)^{-a_{i}a_{j}}\right]\vartheta_{\alpha\beta}^{-1}\left(\sum_{i}a_{i}z_{i}\right)\\
\left\langle \prod_{i}\sigma_{i}\left(z_{i}\right)\right\rangle  & = & \me^{-S_{\text{cl}}}\left|\det W\right|^{-\frac{1}{2}}\vartheta_{1}\left(\theta_{1}z_{1}+\theta_{2}z_{2}-\left(1-\theta_{3}\right)z_{3}\right)^{\frac{1}{2}}\\
 &  & \qquad\qquad\qquad\vartheta_{1}\left(z_{12}\right)^{\frac{1}{2}}\prod_{i<j}^{3}\vartheta_{1}\left(z_{ij}\right)^{-\frac{1}{2}\left(1-\theta_{i}\right)\left(1-\theta_{j}\right)-\frac{1}{2}\theta_{i}\theta_{j}}\\
\left\langle \sigma\left(z_{1}\right)\tau\left(z_{2}\right)\tau^{\prime}\left(z_{3}\right)\right\rangle  & = & \left\langle \prod_{i}\sigma_{i}\left(z_{i}\right)\right\rangle \vartheta_{1}\left(z_{12}\right)^{-\theta_{1}}\vartheta_{1}\left(z_{13}\right)^{\theta_{1}}\vartheta_{1}\left(z_{23}\right)^{\theta_{2}-\theta_{3}-1}\\
 &  & \qquad\qquad\qquad\qquad\qquad\qquad\qquad\qquad\quad\Phi\left(z_{1},z_{2},z_{3}\right)\end{eqnarray*}
\end{minipage}
\tabularnewline
\begin{center}~\\~\\with\\~\end{center}\tabularnewline
\begin{minipage}[c]{1.0\columnwidth}%
\begin{multline*}
\Phi\left(z_{1},z_{2},z_{3}\right)\equiv\Bigg\{\theta_{3}F_{3}\left(z_{3},z_{2}\right)+\bigg[c_{1}\frac{\vartheta_{1}\left(z_{23}\right)}{\vartheta_{1}\left(z_{13}\right)}\vartheta_{1}\left(-z_{1}\left(1-\theta_{1}\right)+z_{2}\theta_{2}+z_{3}\theta_{3}\right)\\
+c_{2}\vartheta_{1}\left(z_{1}\theta_{1}+z_{2}\theta_{2}-z_{3}\left(1-\theta_{3}\right)\right)\bigg]c_{3}\vartheta_{1}\left(z_{1}\theta_{1}-z_{2}\left(1-\theta_{2}\right)+z_{3}\theta_{3}\right)\Bigg\}\end{multline*}
\end{minipage}
~\\~\\~\\~\tabularnewline
\hline
\end{tabular}
\caption{Correlators required for the calculation of $\mathcal{A}$, up to overall normalization factors. The term $F_{3}\left(w,z\right)$
is given by (\ref{eq:extwist_f}).\label{fig:correlators}}
\end{center}}}

Recall that in section \ref{sec:stringcalc}, we found three contributions to $A$, which we denoted $A_{1}$, $A_{2}$ and $A_{3}$. We begin
here with the term $A_{1}$, which we previously argued ought to correspond to a vertex renormalization, limit 1 of figure
\ref{fig:limits}.  Using the correlators given, we find the rather involved expression\begin{multline}
A_{1}=\frac{1}{4}\overline{u}_{1}u_{2}\me^{-S_{\text{cl}}}\left(\sum_{i=1}^{3}\Phi^{i}\left(z_{1},z_{2},z_{3}\right)\right)\\
\vartheta_{1}\left(z_{12}\right)^{k_{1}\cdot k_{2}-\frac{3}{2}}\vartheta_{1}\left(z_{13}\right)^{k_{1}\cdot
k_{3}+\frac{1}{2}\left(\theta_{1}^{1}+\theta_{1}^{2}+\theta_{1}^{3}\right)-1}\vartheta_{1}\left(z_{23}\right)^{k_{2}\cdot
k_{3}+\frac{1}{2}\left(\theta_{2}^{1}+\theta_{2}^{2}+\theta_{2}^{3}\right)-3}\\ \left(\prod_{i=1}^{3}\left|\det
W^{i}\right|^{-\frac{1}{2}}\vartheta_{1}\left(\theta_{1}^{i}z_{1}+\theta_{2}^{i}z_{2}+(\theta_{3}^{i}-1)z_{3}\right)^{\frac{1}{2}}\vartheta_{1}\left(z_{12}\right)^{\frac{1}{2}}\right)\\
\sum_{\alpha\beta}\delta_{\alpha\beta}\vartheta_{\alpha\beta}\left(\tfrac{1}{2}\left(z_{2}-z_{1}\right)\right)\prod_{i=1}^{3}\vartheta_{\alpha\beta}\left(z_{1}q_{1}^{i}+z_{2}\tilde{q}_{2}^{i}+z_{3}\tilde{q}_{3}^{i}\right)\,.\label{eq:quantum_a1}\end{multline}
The function $\Phi^{i}\left(z_{1},z_{2},z_{3}\right)$ in the first line comes from the excited twist correlators, and is displayed
explicitly in figure \ref{fig:correlators}. Factors in the second line come from the combination of all correlators, the third from the
ordinary twist correlation and the fourth from the spin and ghost spin field correlators. The values of
$q_{1}^{i},\,\tilde{q}_{2}^{i},\,\tilde{q}_{3}^{i}$ are given in table \ref{tab:H_charge}.

The phases $\delta_{\alpha\beta}$ may be determined by the requirement that as $z_{1}\rightarrow z_{2}\rightarrow z_{3}$, the amplitude
ought to factor onto the partition function $\mathcal{Z}$ for two D$6$-branes, which contains the term\begin{align} \mathcal{Z} &
\propto\sum_{\alpha\beta}\delta_{\alpha\beta}\vartheta_{\alpha\beta}\left(0\right)^{4}\qquad\left(\delta_{00}=\delta_{11}=+1,\,\delta_{01}=\delta_{10}=-1\right)\,.\label{eq:z_phases}\end{align}
The $H$-charge conservation rule $q_{1}^{i}+\tilde{q}_{2}^{i}+\tilde{q}_{2}^{i}=0$ guarantees that (\ref{eq:quantum_a1}) does indeed have
this property, and so the relative phases in $A_{1}$ must be the same as those in $\mathcal{Z}$. Therefore, we may apply the Riemann
identity (\ref{eq:riemann}), with the result\begin{equation}
A_{1}\propto\vartheta_{1}\left(\tfrac{1}{2}\sum_{i}z_{i}\left(1-\sum_{j}\theta_{i}^{j}\right)\right)\,.\label{eq:a1_afterriemann}\end{equation}

Using the $\mathcal{N}=1$ supersymmetry condition $\sum_{j}\theta_{i}^{j}=1$ and the result $\vartheta_{1}\left(0\right)=0$, we see that
$A_{1}$ always vanishes in supersymmetric models. This appears to support our identification of $A_{1}$ as a vertex renormalization, since
those diagrams vanish in $\mathcal{N}=1$ theories by the non-renormalization theorem.

We now turn to the terms $A_{2}^{\pm}$. Note first that after correlators have been inserted explicitly, the main difference between
(\ref{eq:a2plus}) and (\ref{eq:a2minus}) comes from the spin-dependent parts coming from the external space: $A_{2}^{+}$ contains a factor
of \[
\vartheta_{\alpha\beta}^{-1}\left(\tfrac{1}{2}\left(z_{2}-z_{1}\right)\right)\vartheta_{\alpha\beta}\left(\tfrac{1}{2}\left(z_{2}-z_{1}\right)\right)\vartheta_{\alpha\beta}\left(z_{3}-\tfrac{1}{2}\left(z_{2}+z_{1}\right)\right)\,,\]
whilst $A_{2}^{-}$ contains a factor of \[
\vartheta_{\alpha\beta}^{-1}\left(\tfrac{1}{2}\left(z_{2}-z_{1}\right)\right)\vartheta_{\alpha\beta}\left(-\tfrac{1}{2}\left(z_{2}-z_{1}\right)\right)\vartheta_{\alpha\beta}\left(-z_{3}+\tfrac{1}{2}\left(z_{2}+z_{1}\right)\right)\,.\]
Since the $\vartheta$-functions are odd or even, the difference is superficial and the terms $A_{2}^{\pm}$ may be joined together to give
one single term:\begin{multline} A_{2}=\frac{1}{2}\overline{u}_{1}u_{2}\me^{-S_{\text{cl}}}\left(k_{1}\cdot
k_{3}\frac{\theta_{1}^{\prime}\left(z_{12}\right)}{\theta_{1}\left(z_{12}\right)}+k_{3}\cdot
k_{3}\frac{\theta_{1}^{\prime}\left(z_{23}\right)}{\theta_{1}\left(z_{23}\right)}\right)\\ \vartheta_{1}\left(z_{12}\right)^{k_{1}\cdot
k_{2}-\frac{1}{2}}\vartheta_{1}\left(z_{13}\right)^{k_{1}\cdot
k_{3}+\frac{1}{2}\left(\theta_{1}^{1}+\theta_{1}^{2}+\theta_{1}^{3}\right)-2}\vartheta_{1}\left(z_{23}\right)^{k_{2}\cdot
k_{3}+\frac{1}{2}\left(\theta_{2}^{1}+\theta_{2}^{2}+\theta_{2}^{3}\right)-2}\\ \left(\prod_{i=1}^{3}\left|\det
W^{i}\right|^{-\frac{1}{2}}\vartheta_{1}\left(\theta_{1}^{i}z_{1}+\theta_{2}^{i}z_{2}+\left(\theta_{3}^{i}-1\right)z_{3}\right)^{\frac{1}{2}}\vartheta_{1}\left(z_{12}\right)^{\frac{1}{2}}\right)\\
\sum_{\alpha\beta}\delta_{\alpha\beta}\vartheta_{\alpha\beta}\left(-\tfrac{1}{2}z_{1}-\tfrac{1}{2}z_{2}+z_{3}\right)\prod_{i=1}^{3}\vartheta_{\alpha\beta}\left(\left(\tfrac{1}{2}-\theta_{1}^{i}\right)z_{1}+\left(\tfrac{1}{2}-\theta_{2}^{i}\right)z_{2}-\theta_{3}z_{3}\right)\,.\label{eq:quantum_a2}\end{multline}

In a supersymmetric set-up, $A_{2}$ may be seen to vanish in exactly the same way as $A_{1}$. However, as it has kinematic factors in
front, it does not contribute significantly when the vertex operators are far apart and so we do not associate it with a vertex
renormalization.

More interesting are the terms $A_{3}^{\pm}$, which also join together to give\begin{multline}
A_{3}=-\frac{1}{2}\overline{u}_{1}u_{2}\me^{-S_{\text{cl}}}k_{2}\cdot k_{3}\\ \vartheta_{1}\left(z_{12}\right)^{k_{1}\cdot
k_{2}-\frac{3}{2}}\vartheta_{1}\left(z_{13}\right)^{k_{1}\cdot
k_{3}+\frac{1}{2}\left(\theta_{1}^{1}+\theta_{1}^{2}+\theta_{1}^{3}\right)-1}\vartheta_{1}\left(z_{23}\right)^{k_{2}\cdot
k_{3}+\frac{1}{2}\left(\theta_{2}^{1}+\theta_{2}^{2}+\theta_{2}^{3}\right)-3}\\ \left(\prod_{i=1}^{3}\left|\det
W^{i}\right|^{-\frac{1}{2}}\vartheta_{1}\left(\theta_{1}^{i}z_{1}+\theta_{2}^{i}z_{2}+\left(\theta_{3}^{i}-1\right)z_{3}\right)^{\frac{1}{2}}\vartheta_{1}\left(z_{12}\right)^{\frac{1}{2}}\right)\\
\sum_{\alpha\beta}\delta_{\alpha\beta}\vartheta_{\alpha\beta}\left(-\tfrac{1}{2}z_{1}+\tfrac{3}{2}z_{2}-z_{3}\right)\prod_{i=1}^{3}\vartheta_{\alpha\beta}\left(\left(\tfrac{1}{2}-\theta_{1}^{i}\right)z_{1}+\left(\tfrac{1}{2}-\theta_{2}^{i}\right)z_{2}-\theta_{3}z_{3}\right)\,.\label{eq:quantum_a3}\end{multline}

Again the phases may be determined from factorization on the partition function and so we may apply the Riemann identity, leading
to\begin{multline} A_{3}=-\overline{u}_{1}u_{2}\me^{-S_{\text{cl}}}k_{2}\cdot k_{3}\\ \vartheta_{1}\left(z_{12}\right)^{k_{1}\cdot
k_{2}-\frac{3}{2}}\vartheta_{1}\left(z_{13}\right)^{k_{1}\cdot
k_{3}+\frac{1}{2}\left(\theta_{1}^{1}+\theta_{1}^{2}+\theta_{1}^{3}\right)-1}\vartheta_{1}\left(z_{23}\right)^{k_{2}\cdot
k_{3}+\frac{1}{2}\left(\theta_{2}^{1}+\theta_{2}^{2}+\theta_{2}^{3}\right)-3}\\ \left(\prod_{i=1}^{3}\left|\det
W^{i}\right|^{-\frac{1}{2}}\vartheta_{1}\left(\theta_{1}^{i}z_{1}+\theta_{2}^{i}z_{2}+\left(\theta_{3}^{i}-1\right)z_{3}\right)^{\frac{1}{2}}\vartheta_{1}\left(z_{12}\right)^{\frac{1}{2}}\right)\\
\vartheta_{1}\left(\tfrac{1}{2}\sum_{i}\left(1-\sum_{j}\theta_{i}^{j}\right)z_{i}+z_{2}-z_{3}\right)\prod_{k}\vartheta_{1}\left(\tfrac{1}{2}\sum_{i,j}M_{kj}\theta_{i}^{j}z_{i}+\tfrac{1}{2}\left(z_{2}-z_{1}-z_{3}\right)\right)\,,\label{eq:quantum_a3_afterriemann}\end{multline}
where the matrix $M$ is\[ M_{kj}=\left(\begin{array}{ccc} -1 & 1 & 1\\ 1 & -1 & 1\\ 1 & 1 & -1\end{array}\right)\,.\] This time, the
result does \emph{not} vanish in an $\mathcal{N}=1$ theory, which suggests that we should identify $A_{3}$ as a field
renormalization. Also, since $A_{3}$ has kinematic factors in front, it will be most significant when we take two vertex operators to be
close together; therefore, we confirm the intuitive results of figure \ref{fig:limits}.

As a summary, the three terms that we have found have the following properties:
\begin{center}\begin{tabular}{ccc}
Term& Vanishes by Riemann identity?& Kinematic prefactors?\vspace{0.5ex}\tabularnewline \hline &
\vspace{-1.5ex}&
\tabularnewline
$A_{1}$&
$\surd$&
$\times$\vspace{1ex}\tabularnewline
$A_{2}$&
$\surd$&
$\surd$\vspace{1ex}\tabularnewline
$A_{3}$&
$\times$&
$\surd$\vspace{1ex}\tabularnewline
\end{tabular}\end{center}

It appears that we have a problem if we want an $\mathcal{N}=1$ theory, since the only term which is not killed by a Riemann identity,
$A_{3}$, appears to disappear on-shell via \begin{equation} k_{2}\cdot k_{3}=\frac{1}{2}m^{2}=0\,.\label{eq:quantum_k2k3}\end{equation} In
fact, we shall see in section \ref{sec:beta} that the integral over $y_{i}$ yields poles which should be cancelled with the kinematic
prefactor before going on-shell. In this manner finite answers are obtained. In non-supersymmetric theories, we will see that the
structure of $A_{2}$ and $A_{3}$ (and in particular, the relative minus sign between them) is such that the poles in the combined
expression cancel.

In the case where one of the branes go through an orbifold fixed point, the correlators of spin and twist operators will be modified
\cite{Atick:1988kd}.  However, the differences must only be in the spin-dependent terms, i.e. the last lines of (\ref{eq:quantum_a1}),
(\ref{eq:quantum_a2}) and (\ref{eq:quantum_a3}). In order that the amplitude factor onto the twisted partition function, these
modifications must be exactly those that are made to the spin-dependent terms in the partition function, and so it is possible to write
down the above expressions in the case of orbifold fixed points without repeating the calculation explicitly.  The situation is similar in
the presence of O-planes, where M\"obius strip diagrams may be present.

\section{The classical action\label{sec:classical}}

We now return to the question of how to treat the global monodromy conditions and extract the classical contribution to the action. The
classical action for each pair of complex coordinate can be written\begin{equation}
S_{\text{cl}}=\frac{1}{4\pi\alpha^{\prime}}\int\mathrm{d}^{2}z\,\left(\partial
X_{\text{cl}}\overline{\partial}\overline{X}_{\text{cl}}+\overline{\partial}X_{\text{cl}}\partial\overline{X}_{\text{cl}}\right)\,.\label{eq:class_scl}\end{equation}
The linear decomposition of $X_{\text{cl}}$ can be defined,\begin{equation}
\Delta_{\gamma}X_{\text{cl}}=v_{a}=W_{a}^{i}c_{i}\label{eq:class_mono}\end{equation} where the displacements $v_{a}$ are determined from
the global monodromy conditions (i.e. by comparing the displacement of $X$ under combinations of twists that add up to zero), as discussed
earlier. The coefficients $c_{i}$ are to be determined from them; \begin{equation}
c_{i}=v_{a}(W^{-1})_{i}^{a}\,.\label{eq:class_ci}\end{equation} From the definition of $\Delta X$ we see that we must have\begin{eqnarray}
\partial\overline{X}_{\text{cl}} & = & \sum_{i=1}^{2}c_{i}\omega_{i}\nonumber \\ \partial X_{\text{cl}} & = &
c_{3}\omega_{3}^{\prime}\,.\label{eq:class_dX}\end{eqnarray} Inserting into $S_{cl}$ gives\begin{equation}
S_{\text{cl}}=\frac{1}{4\pi\alpha^{\prime}}\left(\sum_{i,j}^{2}c_{i}c_{j}^{*}I_{ij}+|c_{3}|^{2}I_{33}\right)\label{eq:class_scl2}\end{equation}
where \begin{eqnarray} I_{ij} & = & \int\mathrm{d}^{2}z\,\omega^{i}\bar{\omega}^{j}\nonumber \\ I_{33} & = &
\int\mathrm{d}^{2}z\,|\omega^{3}|\label{eq:class_I}\end{eqnarray} In order to determine the $I_{ij}$ we may perform a canonical dissection
of the torus \cite{Atick:1988kd}. In this case the dissection is as shown in figure \ref{fig:dissection}.

\FIGURE[t]{\parbox{\hsize}{\begin{center} \includegraphics{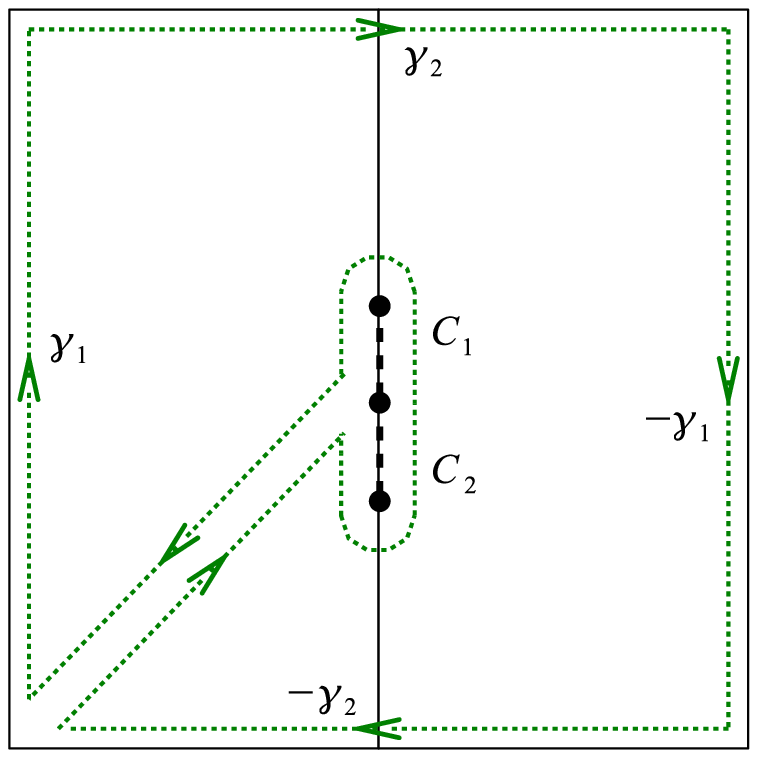}
\caption{The dissection of the torus for three point diagrams. \label{fig:dissection}}
\end{center}}}
In terms of the cycles and two spurs shown in the figure, the $I_{ij}$ are given by \cite{Atick:1988kd}\begin{multline}
I_{ij}=\oint_{\gamma_{1}}\sdif{z}\omega^{i}\oint_{\gamma_{2}}\sdif{\bar{z}}\bar{\omega}^{j}-\oint_{\gamma_{2}}\sdif{z}\omega^{i}\oint_{\gamma_{1}}\sdif{\bar{z}}\bar{\omega}^{j}\\
+\int_{C_{1}}\omega^{i}\int_{C_{2}}\bar{\omega}^{j}+\sum_{l=1}^{2}\frac{1}{\left(1-e^{-2\pi
i\theta_{l}}\right)}\int_{C_{l}}\omega^{i}\int_{C_{l}}\bar{\omega}^{j}\,.\label{eq:class_I2}\end{multline} Evaluating the contour
integrals explicitly we find the relations
\begin{eqnarray}
\int_{C_{1}}\bar{\omega}^{3} & = & \frac{-1}{1-\me^{-2\pi\mi\theta_{2}}}\, W_{3}^{3}\nonumber \\ \int_{C_{2}}\bar{\omega}^{3} & = &
\frac{1}{1-\me^{2\pi\mi\theta_{2}}}\, W_{3}^{3}\nonumber \\ \int_{C_{1}}\omega^{i=1,2} & = & \frac{-1}{1-\me^{-2\pi\mi\theta_{2}}}\,
W_{3}^{i}\nonumber \\ \int_{C_{2}}\omega^{i=1,2} & = & \frac{1}{1-\me^{2\pi\mi\theta_{2}}}\,
W_{3}^{i}\label{eq:class_dissectionintegrals}\end{eqnarray} whence we determine\begin{equation}
I_{ij}=W_{a}^{i}\bar{W}_{b}^{j}K^{ab}\label{eq:class_I3}\end{equation} where \begin{equation} K^{ab}=\left(\begin{array}{ccc} 0 & -\mi &
0\\ \mi & 0 & 0\\ 0 & 0 & \alpha\end{array}\right)\label{eq:class_K}\end{equation} and\begin{equation}
\alpha=\frac{1}{8\,\sin(\pi\theta_{1})\sin(\pi\theta_{2})\sin(\pi\theta_{3})}\,.\label{eq:class_alpha}\end{equation} This gives
\begin{equation} 4\pi\alpha^{\prime}S_{\text{cl}}=\sum_{i,\, j}^{2}c_{i}c_{j}^{*}W_{a}^{i}\overline{W}_{b}^{j}\,
K^{ab}+c_{3}c_{3}^{*}W_{a}^{3}\overline{W}_{b}^{3}K^{ab*}\label{eq:class_scl3}\end{equation} where we have multiplied by an extra factor
of $\frac{1}{2}$ to factor out half the world sheet after the $Z_{2}$ involution. Using the monodromy conditions this can be reduced to
\begin{equation}
4\pi\alpha^{\prime}S_{\text{cl}}=v_{a}v_{b}^{*}K^{ab}-c_{3}W_{a}^{3}v_{b}^{*}K^{ab}-c_{3}^{*}\overline{W}_{b}^{3}v_{a}K^{ab}+2|c_{3}|^{2}W_{a}^{3}\overline{W}_{b}^{3*}K^{ab*}\,.\label{eq:class_scl4}\end{equation}
Inserting $K^{ab}$ yields\begin{multline}
8\pi\alpha^{\prime}S_{\text{cl}}=\alpha\left(|v_{3}-c_{3}W_{3}^{3}|^{2}+|c_{3}W_{3}^{3}|^{2}\right)\\ +\left(\mi\,|c_{3}|^{2}\,
W_{1}^{3}\overline{W}_{2}^{3*}-\mi\,(v_{1}-c_{3}W_{1}^{3})(v_{2}^{*}-c_{3}^{*}\overline{W}_{2}^{3})\,+\,\text{H.C.}\right)\,.\label{eq:class_sclfinal}\end{multline}
This is the main expression for the classical action. Note that we need only determine the coefficient $c_{3}$ explicitly; otherwise, all
that remains to do is to find the $W_{a}^{j}$ in the various limits and insert their values.

\section{Limiting cases of the classical action\label{sec:classical-limits}}

We are now in a position to apply the results derived thus far to the four limiting cases laid out in figure \ref{fig:limits}. We begin by
looking at the classical action in these limits, which we expect to factor onto combinations of the classical contribution to the
partition function,\begin{equation} \mathcal{Z}_{\text{cl}}=\me^{-tY^{2}/2\pi\alpha^{\prime}}\,,\label{eq:zcl}\end{equation} and the
tree-level Yukawa coupling, \begin{equation} \mathcal{Y}_{\text{cl}}=\me^{-\text{Area}/2\pi\alpha^{\prime}}\,.\label{eq:ycl}\end{equation}

\subsection{The partition function limit: $y_{1}\rightarrow y_{2}\rightarrow y_{3}$,
$t\rightarrow\infty$\label{sub:limit0}}

Consider first the factorization of the classical part of the action, in a single $T^{2}$. In this limit, we expect to
find\begin{equation} \me^{-S_{\text{cl}}}\rightarrow\mathcal{Z}_{\text{cl}}\,\mathcal{Y}_{\text{cl}}\,,\label{eq:lim_fac_zy}\end{equation}
which is indeed the case. To see this, one must approximate the $W$ integrals as $y_{1}\rightarrow y_{2}\rightarrow y_{3}$. Such an
approximation is presented in appendix \ref{sub:Wlimit0}, with the result\begin{equation} W=\left(\begin{array}{ccc} \mi t & \mi t & -\mi
t\\ 1 & 1 & 1\\ W_{3}^{1} & W_{3}^{2} & W_{3}^{3}\end{array}\right)\,.\label{eq:text_lim_fac_wmat}\end{equation} Inserting the first two
rows into the monodromy conditions $v_{a}=W_{a}^{i}c_{i}$ gives \begin{eqnarray} \mi t\left(c_{1}+c_{2}-c_{3}\right) & = & v_{1}\nonumber
\\ c_{1}+c_{2}+c_{3} & = & v_{2}\label{eq:lim_fac_mono2}\end{eqnarray} with solution\begin{eqnarray} c_{3} & = &
\frac{v_{2}}{2}-\frac{v_{1}}{2\mi t}\,.\label{eq:lim_fac_c3}\end{eqnarray} Since $W_{3}^{3}$ is vanishingly small in the limit
(c.f. eq. \ref{eq:lim_fac_W33}), we may also write\begin{equation} W_{3}^{3}c_{3}=0\,.\label{eq:lim_fac_W33c3}\end{equation} Inserting
these results into (\ref{eq:class_sclfinal}) leads to the action\begin{equation}
2\pi\alpha^{\prime}S_{\text{cl}}=\frac{\alpha}{4}|v_{3}|^{2}+t\left|\frac{v_{2}}{2}\right|^{2}+\frac{\left|v_{1}\right|^{2}}{4t}\,.\label{eq:lim_fac_scl}\end{equation}

Values must be inserted into this expression for the physical displacements $v_{i}$. First, note that $v_{3}$ is given by the Pochhammer
contour\begin{equation}
v_{3}=4\sin(\pi\theta_{1})\sin(\pi\theta_{2})e^{\mi\pi\left(\theta_{2}-\theta_{1}\right)}f_{12}\,,\label{eq:lim_fac_poch}\end{equation}
where $f_{12}$ is the spacetime displacement between vertices one and two. Hence, the first term in the action is the area of a
triangle\begin{equation}
\frac{\alpha}{4}|v_{3}|^{2}=\frac{\sin(\pi\theta_{1})\sin(\pi\theta_{2})|f_{12}|^{2}}{2\sin(\pi\theta_{3})}\,,\label{eq:lim_fac_triangle}\end{equation}
and is the classical part of the tree level Yukawa coupling.

With our definitions, $\left|v_{2}\right|$ is twice the inter-brane separation in a given torus, $v_{2}=2\mi Y_{i}$, since the contour
integral goes across twice the fundamental domain of the annulus (c.f.  figure \ref{fig:contours}). The second term in
(\ref{eq:lim_fac_scl}) is thus the partition function term $Y^{2}t/2\pi\alpha^{\prime}$, seen in eq. \ref{eq:zcl}.

From figure \ref{fig:annulus} we see that the displacement $v_{1}$ may be taken to zero. In a compact space, it should also be summed over
all wrappings; \begin{equation} v_{1}=2\pi Ln\,,\label{eq:lim_fac_v1resum}\end{equation} where $2\pi L$ is the wrapping length of the
$v_{1}$ brane. Depending on the range of $t$ under consideration, it may be appropriate to apply the Poisson resummation
formula,\begin{equation} \sum_{n=-\infty}^{\infty}\me^{-\pi an^{2}+2\pi\mi
bn}=a^{-\frac{1}{2}}\sum_{m=-\infty}^{\infty}\me^{-\pi\left(m-b\right)^{2}/a}\,,\label{eq:poisson}\end{equation} after which we see that
for a single torus\begin{align} e^{-S_{\text{cl}}} &
=\frac{\sqrt{2\alpha^{\prime}t}}{L}\me^{-\text{Area}/2\pi\alpha^{\prime}}\me^{-tY^{2}/2\pi\alpha^{\prime}}\sum_{m=-\infty}^{\infty}e^{-2\pi\alpha^{\prime}tm^{2}/L}\,.\label{eq:lim_fac_escl_resummed}\end{align}
The extra factors of $\sqrt{t}$ here are due to Kaluza-Klein modes propagating in the loop. The condition for the resummation to be valid
is that $2\pi\alpha^{\prime}t>L$, and so we will get a factor of $\sqrt{t}$ appearing in the amplitude for each torus where
$2\pi\alpha^{\prime}t>L_{i}$.  We discuss the consequences of this in section \ref{sec:beta} below.

\subsection{The vertex correction limit: $\left(t-y_{3}\right)\rightarrow\infty$,
generic $y_{i}$\label{sec:limit1}}

A derivation of the $W$ integrals in this limit is presented in Appendix \ref{sub:Wlimit1}, where we show that up to terms which are
suppressed by powers of $t$,\begin{equation} W=\left(\begin{array}{ccc} \mi\Delta t+D & AC & -\mi\Delta t+D+A\\ 1 & 1 & 1\\ B & BC &
B\end{array}\right)\,,\label{eq:text_lim_gen_wmat}\end{equation} with\begin{eqnarray} A & = &
-\frac{\mi}{2}\left[\cot\left(\pi\theta_{1}\right)+\cot\left(\pi\theta_{3}\right)\right]\,=\,-\frac{i}{2}\left(\frac{\sin(\pi\theta_{2})}{\sin(\pi\theta_{1})\sin(\pi\theta_{3})}\right)\nonumber
\\ B & = & -2\mi e^{i\pi\theta_{2}}\sin(\pi\theta_{2})\nonumber \\ C & = &
-\frac{\Gamma(\theta_{2})}{\Gamma(1-\theta_{1})\Gamma(1-\theta_{3})}e^{2\pi y_{2}\theta_{1}}\,,\nonumber \\ D & = &
-\frac{\mi}{2\pi}\left[2\gamma_{E}+\psi\left(1-\theta_{1}\right)+\psi\left(1-\theta_{3}\right)\right]\label{eq:text_lim_gen_funcs}\end{eqnarray}
and $\Delta t=t-y_{3}$.

As the nine elements of $W$ are expressible in terms of five independent quantities, we can see that there will be some over-determination
in the monodromy conditions. That is, the equation\begin{equation} v_{a}=W_{a}^{i}c_{i}\label{eq:lim_gen_vrel}\end{equation} will yield
relations between the $v_{a}$. But recall that the $v_{a}$ are physical displacements, relating the sides of triangles to their heights,
and consequently the $W_{a}^{i}$ matrix we have determined ought to give displacements corresponding to the actual Yukawa triangles.  We
will see that it does presently.

Notice first that since $W_{1}^{2}$ and $W_{3}^{2}$ contain the factor $e^{2\pi y_{2}\theta_{1}}$, they are exponentially dominant in the
$v_{1}$ and $v_{3}$ conditions and so we must have $c_{1},\, c_{3}\gg c_{2}$.  The $v_{2}$ condition is therefore \begin{equation}
c_{1}+c_{3}\simeq v_{2}\,.\label{eq:lim_gen_cdom}\end{equation} Using this result together with the conditions for $v_{3}$ and $v_{1}$, we
find\begin{alignat}{1} c_{3} & =\frac{1}{2\mi\Delta t-A}\left(\left(\mi\Delta t+D-A\right)v_{2}-v_{1}+\frac{A}{B}v_{3}\right)\nonumber \\
& \simeq\frac{v_{2}}{2}+\frac{\mi}{2\Delta t}\left(v_{1}-\frac{A}{B}v_{3}\right)\,.\label{eq:lim_gen_c3}\end{alignat} To obtain the full
$S_{cl}$ via (\ref{eq:class_scl}), we will need
\begin{equation}
c_{3}W_{3}^{3}=c_{3}B=\frac{\hat{v}_{2}}{2}+\frac{\mi A}{2\Delta t}\left(\hat{v}_{1}-v_{3}\right)\,,\label{eq:lim_gen_c3W3}\end{equation}
where \begin{eqnarray} \hat{v}_{1} & = & \frac{B}{A}v_{1}\nonumber \\ \hat{v}_{2} & = &
Bv_{2}\,.\label{eq:lim_gen_v1hatv2hat}\end{eqnarray} At this point it is appropriate to say something about the geometry of the
situation. Notice that as $A$ is pure imaginary, (\ref{eq:lim_gen_c3W3}) shows that both $\hat{v}_{1}$ and $\hat{v}_{2}$ must have the
same phase as $v_{3}$; arranging things so that $v_{1}$ is purely real and $v_{2}$ pure imaginary, $v_{3}$ must have an overall phase of
$\me^{\mi\pi\theta_{2}}$. However, since $v_{3}$ is also\begin{equation}
v_{3}=4\sin(\pi\theta_{1})\sin(\pi\theta_{2})e^{i\pi\left(\theta_{2}-\theta_{1}\right)}f_{12}\,,\label{eq:lim_gen_poch}\end{equation} from
the Pochhammer contour, $f_{12}$ must have a phase of $\me^{\mi\pi\theta_{1}}$.  We may therefore define two additional vectors for
$\hat{v}_{1}$ and $\hat{v}_{2}$, which we call $f_{1}$ and $f_{2}$ respectively, such that \begin{eqnarray} \hat{v}_{1} & = &
4\sin\left(\pi\theta_{1}\right)\sin\left(\pi\theta_{2}\right)e^{\mi\pi\left(\theta_{2}-\theta_{1}\right)}\, f_{1}\nonumber \\ \hat{v}_{2}
& = & 4\sin\left(\pi\theta_{1}\right)\sin\left(\pi\theta_{2}\right)e^{\mi\pi\left(\theta_{2}-\theta_{1}\right)}\,
f_{2}\,.\label{eq:lim_gen_v1hatv2hat_2}\end{eqnarray}

\FIGURE[t]{\scalebox{0.95}{\includegraphics{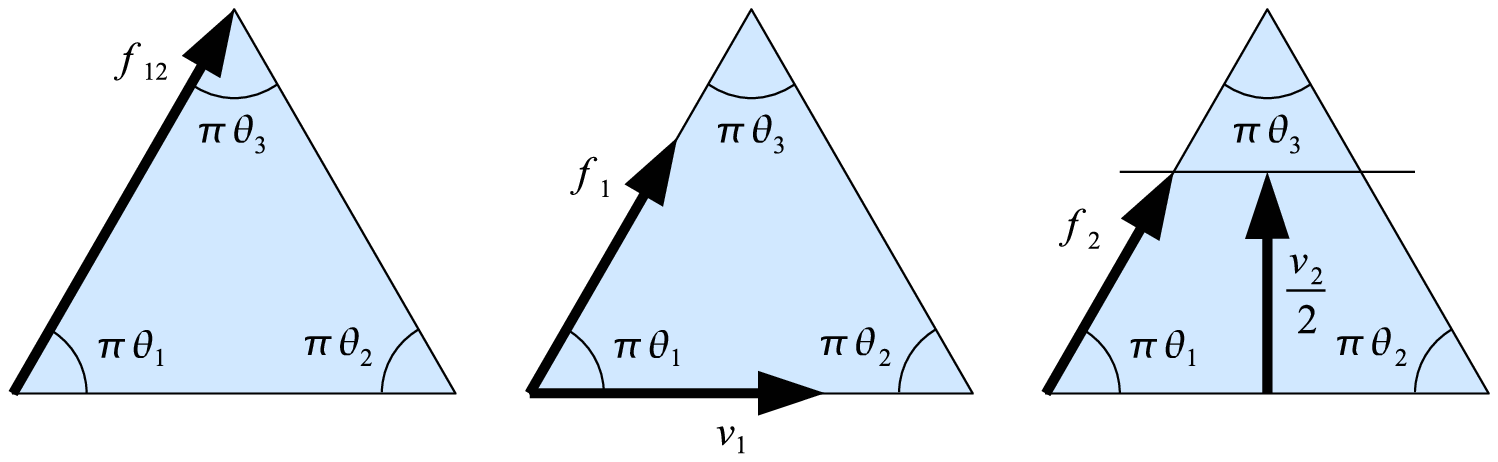}}
\caption{Physical vectors which appear in the global monodromy conditions $W_{a}^{i}c_{i}=v_{a}$.\label{fig:f1f2}} } With these
definitions, $f_{1}$ and $f_{2}$ are projections of $v_{1}$ and $v_{2}/2$ along $f_{12}$:
\begin{align}
f_{1} & =\me^{\mi\pi\theta_{1}}\frac{\sin\left(\pi\theta_{3}\right)v_{1}}{\sin\left(\pi\theta_{2}\right)}\nonumber \\
f_{2} & =\me^{\mi\pi\left(\theta_{1}-\frac{1}{2}\right)}\frac{v_{2}/2}{\sin\left(\pi\theta_{1}\right)}\,.\label{eq:lim_gen_f1f2}
\end{align}
The set-up is illustrated in figure \ref{fig:f1f2}.

Inserting (\ref{eq:lim_gen_c3}, \ref{eq:lim_gen_c3W3}) into the result (\ref{eq:class_scl}) and using $\alpha=\mi A/\left|B\right|^{2}$
gives the classical action as\begin{equation} 2\pi\alpha^{\prime}S_{\text{cl}}=\left|\frac{v_{2}}{2}\right|^{2}\left(\Delta t+\mi
D\right)+\frac{\alpha}{4}\left|v_{3}-\hat{v}_{2}\right|^{2}+\frac{\alpha}{8}\left|\hat{v}_{2}\right|^{2}+\mathcal{O}\left(\frac{1}{\Delta
t}\right)\,.\label{eq:lim_gen_scl}\end{equation} The first term in $S_{\text{cl}}$ is the usual inter-brane suppression, $Y^{2}\Delta
t/2\pi\alpha^{\prime}$, showing that small $v_{2}=2\mi Y$ is favoured. The second (and non-standard) term in
$S_{\text{cl}}$,\begin{equation}
\frac{Y^{2}}{2\pi^{2}\alpha^{\prime}}\left[2\gamma_{E}+\psi\left(1-\theta_{1}\right)+\psi\left(1-\theta_{3}\right)\right]\,,\label{eq:lim_gen_funnybit}\end{equation}
favours small values of $\theta_{1}$ and $\theta_{3}$, tending to zero as $\theta_{1}\rightarrow0$, $\theta_{3}\rightarrow0$; this is just
another reflection of the desire of the string not to stretch.

As $v_{2}\rightarrow0$, the third term in $S_{\text{cl}}$ become the area of the large triangle (c.f. equation \ref{eq:lim_fac_triangle}),
and gives the tree level Yukawa coupling. The third term in the action is then the familiar effect of the propagation of a heavy
(generation changing) mode of length $v_{2}/2$ stretched between two branes.  This situation is shown in figure \ref{fig:tree-yuk}. The
geometric interpretation of the $\frac{\alpha}{8}\left|\hat{v}_{2}\right|^{2}$ term is not obvious.

\FIGURE[t]{\parbox{\hsize}{\begin{center} \includegraphics{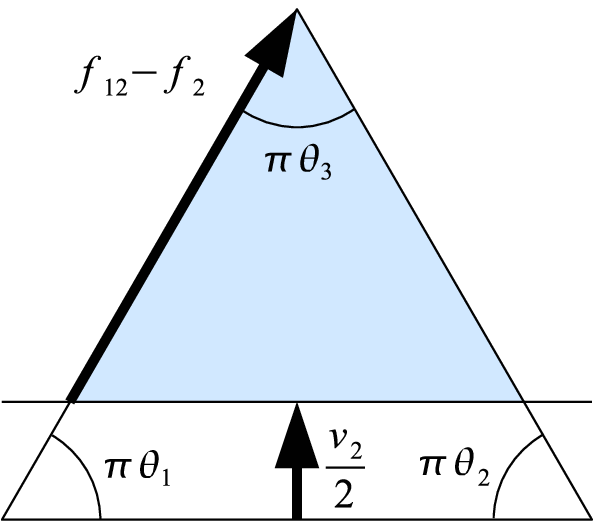}
\caption{As $v_{2}\rightarrow0$, we recover the tree-level Yukawa coupling.\label{fig:tree-yuk}}
\end{center}}}
As $v_{2}\rightarrow0$, we expect to find Kaluza-Klein modes propagating in the loop. This may be seen by retaining the
$\mathcal{O}\left(1/\Delta t\right)$ terms in $S_{\text{cl}}$, which when $v_{2}=0$ reads \begin{equation}
2\pi\alpha'S_{cl}=\frac{\alpha}{4}|v_{3}|^{2}+\frac{\left|v_{1}-\tilde{v}_{3}\right|^{2}}{4\Delta t}+\mathcal{O}\left(\frac{1}{\Delta
t^{2}}\right)\,,\label{eq:lim_gen_scl2}\end{equation} where \begin{equation}
\tilde{v}_{3}=\frac{A}{B}v_{3}=e^{-i\pi\theta_{1}}\frac{\sin\left(\pi\theta_{2}\right)}{\sin\left(\pi\theta_{3}\right)}f_{12}\label{eq:lim_gen_v3t}\end{equation}
is the projection of $v_{3}$ along $v_{1}$. The displacement $v_{1}$ is allowed to wrap the brane between the 13 vertices, and we should
sum over all wrappings. From figure \ref{fig:contours}, we see that $v_{1}$ must be at least as large as $\tilde{v}_{3}$ and so it is
convenient to let \begin{equation} v_{1}=2\pi Ln+\tilde{v}_{3}\,,\label{eq:lim_gen_v1resum}\end{equation} after which the resummation
proceeds just as in sec. \ref{sub:limit0}, with the result

\begin{align}
e^{-S_{\text{cl}}} & =\frac{\sqrt{2\alpha^{\prime}\Delta t}}{L}\me^{-\text{Area}/2\pi\alpha^{\prime}}\me^{-\Delta
tY^{2}/2\pi\alpha^{\prime}}\sum_{m=-\infty}^{\infty}e^{-2\pi\alpha^{\prime}\Delta tm^{2}/L}\,.\label{eq:lim_gen_escl_resummed}\end{align}

\subsection{The purely twisted loop: $\left(t-y_{3}\right)\rightarrow0$, generic
$y_{i}$\label{sec:limit2}}

As explained in Appendix \ref{sub:Wlimit2}, we now have\begin{equation} W=\left(\begin{array}{ccc} A+i\Delta t_{1}^{\prime} & AC &
A-i\Delta t_{3}^{\prime}\\ 1 & 1 & 1\\ B & BC & B\end{array}\right)\label{eq:text_lim_twist_wmat}\end{equation} with $A$, \textbf{$B$} and
$C$ defined as per the previous section, and\begin{eqnarray} \Delta t_{1}^{\prime} & = &
\frac{\mi}{2}\left[\frac{\pi^{\theta_{1}-1}}{\theta_{1}}\left(t-y_{3}\right)^{\theta_{1}}+\frac{\pi^{\theta_{3}-1}}{\theta_{3}}\left(t-y_{3}\right)^{\theta_{3}}\right]\nonumber
\\ \Delta t_{3}^{\prime} & = &
\frac{\mi}{2}\left[\frac{\pi^{-\theta_{1}}}{1-\theta_{1}}\left(t-y_{3}\right)^{1-\theta_{1}}+\frac{\pi^{-\theta_{3}}}{1-\theta_{3}}\left(t-y_{3}\right)^{1-\theta_{3}}\right]\,.\label{eq:text_lim_twist_deltats}\end{eqnarray}
The same argument as in section \ref{sec:limit1} then leads to\begin{equation} c_{3}=\frac{v_{2}}{1+\Delta t_{3}^{\prime}/\Delta
t_{1}^{\prime}}+\frac{\mi}{\Delta t_{1}^{\prime}+\Delta
t_{3}^{\prime}}\left(v_{1}-\frac{A}{B}v_{3}\right).\label{eq:lim_twist_c3}\end{equation} For $\theta_{1}<\frac{1}{2}$ and
$\theta_{3}<\frac{1}{2}$, $\lim_{(t-y_{3})\rightarrow0}\Delta t_{3}^{\prime}/\Delta t_{1}^{\prime}=0$, and so the first term remains
finite in the limit. The second term (and therefore $S_{\text{cl}}$) tends to infinity in the limit, unless
\begin{equation}
v_{1}=\frac{A}{B}v_{3}=\tilde{v}_{3}\,,\label{eq:lim_twist_con}\end{equation} in which case $c_{3}=v_{2}$. Note that \begin{equation}
\tilde{v}_{3}=e^{-i\pi\theta_{1}}\frac{\sin\pi\theta_{2}}{\sin\pi\theta_{3}}f_{12}\label{eq:lim_twist_con2}\end{equation} is the
projection of $f_{12}$ along the $v_{1}$ direction, so this condition amounts to constraining our wrapping to triangles which are
congruent to the original. As in the previous section, the origin of this constraint on $v_{1}$ lies in figure \ref{fig:contours}; in
general, $v_{1}$ must be at least as large as $\tilde{v}_{3}$, but in the limit where we take vertex operators 1 and 3 to opposite sides
of the worldsheet $v_{1}$ is pinched into a limit, leading to (\ref{eq:lim_twist_con}).

Substituting in to (\ref{eq:class_sclfinal}), one finds the classical action\begin{equation}
2\pi\alpha'S_{\text{cl}}=2\left|\frac{v_{2}}{2}\right|^{2}\Delta
t_{3}^{\prime}+\frac{\alpha}{4}\left|v_{3}-\hat{v}_{2}\right|^{2}+\frac{3\alpha}{4}\left|\hat{v}_{2}\right|^{2}\,.\label{eq:lim_twist_scl}\end{equation}
The first term, which goes to zero in the limit, is the stretching term already encountered. The second and third terms are area factors
which give various combinations of Yukawa triangles. The three Yukawa couplings in the first diagram of limit 2 of figure \ref{fig:limits}
may be recovered by taking $v_{3}=\hat{v}_{2}$. There are no terms suitable for Poisson resummation in this limit and so, correctly,
Kaluza-Klein modes are not present.

\subsection{The untwisted \& twisted loop limit: $\left(t-y_{3}\right)\rightarrow\infty$,
$y_{2}\rightarrow y_{1}$\label{sec:limit3}}

From appendix \ref{sub:Wlimit3}, the $W$ matrix in this limit is\begin{equation} W=\left(\begin{array}{ccc} \mi\Delta t+E & \mi\Delta t+E
& -\mi\Delta t-E\\ 1 & 1 & 1\\ W_{3}^{1} & W_{3}^{2} & W_{3}^{3}\end{array}\right)\label{eq:lim_untwist_wmat}\end{equation}
where\begin{equation}
E=-\frac{\mi}{2\pi}\left(2\gamma_{E}+\psi\left(\theta_{3}\right)+\psi\left(1-\theta_{3}\right)\right)\,.\label{eq:lim_untwist_E}\end{equation}
The monodromy conditions $v_{a}=W_{a}^{i}c_{i}$ now lead to\begin{equation} c_{3}=\frac{v_{2}}{2}-\frac{v_{1}}{2\left(\mi\Delta
t-E\right)}\,.\label{eq:lim_untwist_c3}\end{equation}

In this limit, $W_{3}^{i}\rightarrow0$ (c.f. eq. \ref{eq:lim_untwist_W3i}).  Therefore, the action is\begin{equation}
2\pi\alpha'S_{cl}=\left|\frac{v_{2}}{2}\right|^{2}\left(\Delta t-\mi
E\right)+\frac{\alpha}{4}\left|v_{3}\right|^{2}+\frac{\left|v_{1}\right|^{2}}{4\left(\Delta t-\mi
E\right)}\,.\label{eq:lim_untwist_scl}\end{equation} The first specimen is the standard string stretching suppression.  The term in $-\mi
E$ has a minimum around $\theta_{3}=\frac{1}{2}$ -- like (\ref{eq:lim_gen_funnybit}), it is also attempting to prevent the string
stretching. The third term is the tree level Yukawa coupling; as we see from figure \ref{fig:limits}, no other couplings are present in
this diagram. The fourth term may be Poisson resummed, after which the action takes the form\begin{align} e^{-S_{\text{cl}}} &
=\frac{\sqrt{2\alpha^{\prime}\left(\Delta t-\mi E\right)}}{L}\me^{-\text{Area}/2\pi\alpha^{\prime}}\me^{-\left(\Delta t-\mi
E\right)Y^{2}/2\pi\alpha^{\prime}}\sum_{m=-\infty}^{\infty}e^{-2\pi\alpha^{\prime}\left(\Delta t-\mi
E\right)m^{2}/L}\,.\label{eq:lim_untwist_escl_resummed}\end{align}

\section{Extraction of $\beta$-functions\label{sec:beta}}

To complete our analysis, we will demonstrate how $\beta$-functions may be extracted by considering limit 3 -- that is, the limit with
both twisted and untwisted states in the loop. We concentrate on the field renormalization terms $A_{2}$ and $A_{3}$.

\subsection{Factorization onto partition function\label{sub:qfac}}

Recall that the amplitude takes the form\begin{equation}
\mathcal{A}=\int_{0}^{\infty}\frac{\mathrm{d}t}{t}f\left(t\right)\int\sdif{z_{1}\mathrm{d}z_{2}\mathrm{d}z_{3}}A\left(z_{1},z_{2},z_{3}\right)\,.\label{eq:beta_amp}\end{equation}
The first thing that we must do is determine the overall normalization $f\left(t\right)$ in (\ref{eq:beta_amp}). This may be done by the
same mechanism used to determine phases previously in \ref{sec:quantum}: factorization of the amplitude in the limit where the vertex
operators come together (limit 0 of figure \ref{fig:limits}).

In the factorization limit one expects to find $\mathcal{A}\rightarrow\mathcal{Z\, Y}$.  It was shown in section \ref{sub:limit0} that the
classical part of the amplitude factorizes correctly, so in what follows we concentrate on the quantum piece. For the case of D$6$-branes
in a flat space, the quantum part of the partition function is, in the $t\rightarrow\infty$ limit,\begin{equation}
\mathcal{Z}_{\text{qu}}=\int_{0}^{\infty}\frac{\mathrm{d}t}{4t}t^{-\frac{1}{2}\left(6+1\right)}\me^{\pi
t}\sum_{\alpha\beta}\delta_{\alpha\beta}\vartheta_{\alpha\beta}\left(0\right)^{4}\,,\label{eq:qfac_z}\end{equation} where we have ignored
an infinite volume factor. We will also need the quantum normalization of the tree-level Yukawa coupling, which is
\cite{Burwick:1990tu,Stieberger:1992bj,Erler:1992gt,Stieberger:1992vb,Cvetic:2003ch,Abel:2003vv,Lust:2004cx}\begin{equation}
\mathcal{Y}_{\text{qu}}=\prod_{i=1}^{3}\left[\frac{\Gamma\left(\theta_{1}^{i}\right)\Gamma\left(\theta_{2}^{i}\right)\Gamma\left(\theta_{3}^{i}\right)}{\Gamma\left(1-\theta_{1}^{i}\right)\Gamma\left(1-\theta_{2}^{i}\right)\Gamma\left(1-\theta_{3}^{i}\right)}\right]^{\frac{1}{4}}\,.\label{eq:qfac_y}\end{equation}
We now have nearly all the information required to write down the $A_{i}$ terms in the $t\rightarrow\infty$ limit, save for one
technicality: when we first began to construct correlators in terms of $\vartheta_{1}$-functions in section \ref{sub:corr_bosonic}, we
mentioned that they ought to contain a contribution $r\left(t\right)$ which would regulate the correlators in the limit
$z\rightarrow0$. As we are now interested in the $t$-dependence of the correlators, we can no longer ignore this piece, which simply
corresponds to making the replacement
$\vartheta_{1}\left(z\right)\rightarrow\vartheta_{1}\left(z\right)/\vartheta_{1}^{\prime}\left(0\right)$ everywhere, so that in the
$t\rightarrow\infty$ limit the relevant expansion is just

\begin{equation}
\frac{\vartheta_{1}\left(\mi y\right)}{\vartheta_{1}^{\prime}\left(0\right)}\simeq\frac{\mi\sinh\left(\pi
y\right)}{\pi}\,.\label{eq:qfac_thetarep}\end{equation} No such replacement is made in the spin-structure dependent portion, as it does
not originate from a physical propagator.

As expained in section \ref{sec:quantum}, each of the terms $A_{1},\, A_{2},\, A_{3}$ factors correctly onto the partition function. We
could therefore obtain $f\left(t\right)$ from any combination of these terms; we choose the field renormalization limits $A_{2}$ and
$A_{3}$, as it is a field renormalization we will be ultimately interested in.  To write down $A_{2}$ and $A_{3}$ in the limit, a result
for $\det W$ as $y_{3}\rightarrow y_{2}\rightarrow0$ is required. This is derived in appendix \ref{sub:Wlimit0} as\begin{equation}
\left|\det
W\right|=\frac{8\pi^{2}t}{\Gamma\left(1-\theta_{1}\right)\Gamma\left(1-\theta_{2}\right)\Gamma\left(1-\theta_{3}\right)}y_{12}^{-\theta_{3}}y_{13}^{-\theta_{2}}y_{23}^{-\theta_{1}}y_{2}\left(\theta_{1}y_{1}+\theta_{2}y_{2}-\left(1-\theta_{3}\right)y_{3}\right)\,.\label{eq:qfac_detW}\end{equation}
Notice that the factor of $y_{2}\left(\theta_{2}^{i}y_{2}-\left(1-\theta_{3}^{i}\right)y_{3}\right)$ is exactly that required to make the
twist correlator (\ref{eq:twist_zqu}) symmetric in the $y_{i}$ -- which of course it should be, as it it does not distinguish between
fermions and bosons.

The integrals are simplified if we use translational invariance on the worldsheet to fix $y_{1}=0$ (inserting a factor of $t$ to
compensate) and make the change of variables \begin{equation}
\rho=\frac{y_{2}}{y_{3}}\qquad\qquad\lambda=\frac{t-y_{3}}{t}\,,\label{eq:qfac_cov}\end{equation} so that the integration over all vertex
operator positions (\ref{eq:amp}) is performed as\begin{align} \mathcal{A}_{2}+\mathcal{A}_{3} &
=-\mi\int_{0}^{\infty}\frac{\mathrm{d}t}{t}f\left(t\right)t^{3}\int_{0}^{1}\sdif{\rho}\int_{0}^{1}\sdif{\lambda}\left(A_{2}+A_{3}\right)\,.\label{eq:qfac_amp}\end{align}

Inserting (\ref{eq:qfac_thetarep}) and (\ref{eq:qfac_detW}) into (\ref{eq:quantum_a2}) and (\ref{eq:quantum_a3}), one
finds\begin{multline}
\mathcal{A}_{2}=\int_{0}^{\infty}\frac{\mathrm{d}t}{t}f\left(t\right)g\left(t\right)\sum_{\alpha\beta}\delta_{\alpha\beta}\vartheta_{\alpha\beta}\left(0\right)^{4}\\
\left[k_{1}\cdot k_{3}\int_{0}^{1}\sdif{\rho}\rho^{-1+k_{1}\cdot k_{2}+\frac{1}{2}m^{2}}\left(1-\rho\right)^{-1+k_{2}\cdot
k_{3}-\frac{1}{2}m^{2}}\int_{0}^{1}\sdif{\lambda}\left(1-\lambda\right)^{-1+k_{1}\cdot k_{2}+k_{1}\cdot k_{3}+k_{2}\cdot
k_{3}-\frac{1}{2}m^{2}}\right.\\ \left.+k_{3}^{2}\int_{0}^{1}\sdif{\rho}\rho^{k_{1}\cdot
k_{2}+\frac{1}{2}m^{2}}\left(1-\rho\right)^{-2+k_{2}\cdot
k_{3}-\frac{1}{2}m^{2}}\int_{0}^{1}\sdif{\lambda}\left(1-\lambda\right)^{-2+k_{1}\cdot k_{2}+k_{1}\cdot k_{3}+k_{2}\cdot
k_{3}-\frac{1}{2}m^{2}}\right]\label{eq:qfac_a2}\end{multline} and\begin{multline}
\mathcal{A}_{3}=-\int_{0}^{\infty}\frac{\mathrm{d}t}{t}f\left(t\right)g\left(t\right)\sum_{\alpha\beta}\delta_{\alpha\beta}\vartheta_{\alpha\beta}\left(0\right)^{4}\\
k_{2}\cdot k_{3}\int_{0}^{1}\sdif{\rho}\rho^{-1+k_{1}\cdot k_{2}+\frac{1}{2}m^{2}}\left(1-\rho\right)^{-2+k_{2}\cdot
k_{3}-\frac{1}{2}m^{2}}\int_{0}^{1}\sdif{\lambda}\left(1-\lambda\right)^{-2+k_{1}\cdot k_{2}+k_{1}\cdot k_{3}+k_{2}\cdot
k_{3}-\frac{1}{2}m^{2}}\label{eq:qfac_a3}\end{multline} where

\begin{equation}
g\left(t\right)=t^{-\frac{3}{2}+k_{1}\cdot k_{2}+k_{1}\cdot k_{3}+k_{2}\cdot
k_{3}-\frac{1}{2}m^{2}}\prod_{i=1}^{3}\left[\frac{\Gamma\left(1-\theta_{1}^{i}\right)\Gamma\left(1-\theta_{2}^{i}\right)\Gamma\left(1-\theta_{3}^{i}\right)}{8\pi^{2}}\right]^{\frac{1}{2}}\,,\label{eq:qfac_g}\end{equation}
and an irrelevant overall phase has been dropped. All explicit angular dependence in the exponents of (\ref{eq:qfac_a2},~\ref{eq:qfac_a3})
has vanished via the relation $m^{2}=\sum_{i}\theta_{3}^{i}-1$.

On shell, the momenta obey $k_{1}^{2}=0,\, k_{2}^{2}=0,\, k_{3}^{2}=-m^{2}$, and so\begin{equation} k_{1}\cdot
k_{2}=-\frac{1}{2}m^{2}\,,\qquad k_{1}\cdot k_{3}=\frac{1}{2}m^{2}\,,\qquad k_{2}\cdot
k_{3}=\frac{1}{2}m^{2}\,.\label{eq:qfac_onshell}\end{equation} It is therefore clear that the integrals over $\rho$ contain poles
on-shell. We deal with these by writing the integrals as gamma functions which may then be expanded about the poles: in $A_{3}$, for
instance,\begin{align} \int_{0}^{1}\sdif{\rho}\rho^{-1+k_{1}\cdot k_{2}+\frac{1}{2}m^{2}}\left(1-\rho\right)^{-2+k_{2}\cdot
k_{3}-\frac{1}{2}m^{2}} & =\frac{\Gamma\left(k_{1}\cdot k_{2}+\frac{1}{2}m^{2}\right)\Gamma\left(k_{2}\cdot
k_{3}-\frac{1}{2}m^{2}-1\right)}{\Gamma\left(k_{1}\cdot k_{2}+k_{2}\cdot k_{3}-1\right)}\nonumber \\ & \simeq\frac{\left(k_{1}\cdot
k_{2}+k_{2}\cdot k_{3}\right)}{\left(k_{1}\cdot k_{2}+\frac{1}{2}m^{2}\right)\left(k_{2}\cdot
k_{3}-\frac{1}{2}m^{2}\right)}\,.\label{eq:qfac_a3rig}\end{align} Performing a similar expansion in $A_{2}$ and taking everything on-shell
that does not lead to a pole, we end up with\begin{equation}
\mathcal{A}_{2}+\mathcal{A}_{3}=\int_{0}^{\infty}\frac{\mathrm{d}t}{t}f\left(t\right)g\left(t\right)\sum_{\alpha\beta}\delta_{\alpha\beta}\vartheta_{\alpha\beta}\left(0\right)^{4}\frac{k_{1}\cdot
k_{2}+k_{2}\cdot k_{3}}{k_{1}\cdot k_{2}+\frac{1}{2}m^{2}}\left[\frac{k_{2}\cdot k_{3}-k_{1}\cdot k_{3}}{k_{2}\cdot
k_{3}-\frac{1}{2}m^{2}}-m^{2}\right]\,.\label{eq:qfac_a2a3_poles}\end{equation} The $k_{2}\cdot k_{3}-k_{1}\cdot k_{3}$ pole occurs
because of the relative minus sign between the $A_{2}$ and $A_{3}$ terms, which may be traced back to the results of the picture-changing
in equations (\ref{eq:a2plus}-\ref{eq:a3minus}). The poles in the last two terms cancel when we go on-shell, yielding the finite
result\begin{align} \mathcal{A}_{2}+\mathcal{A}_{3} &
=\left(1-m^{2}\right)\int_{0}^{\infty}\frac{\mathrm{d}t}{t}f\left(t\right)g\left(t\right)\sum_{\alpha\beta}\delta_{\alpha\beta}\vartheta_{\alpha\beta}\left(0\right)^{4}\,.\label{eq:qfac_a2a3_finite}\end{align}

Comparison of (\ref{eq:qfac_a2a3_finite}) with the partition function (\ref{eq:qfac_z}) and Yukawa normalization (\ref{eq:qfac_y}) then
shows that\begin{equation} f\left(t\right)=t^{-2}\me^{\pi
t}\frac{1}{1-m^{2}}\mathcal{Y}_{\text{qu}}\prod_{i=1}^{3}\left[\frac{\Gamma\left(1-\theta_{1}^{i}\right)\Gamma\left(1-\theta_{2}^{i}\right)\Gamma\left(1-\theta_{3}^{i}\right)}{8\pi^{2}}\right]^{-\frac{1}{2}}\,.\label{eq:qfac_ft}\end{equation}
In a compact space, $\mathcal{Z}_{\text{qu}}$ is modified by a power of $t^{\frac{3}{2}}$ as $t\rightarrow\infty$, since the Kaluza-Klein
modes are quenched in this limit and ought to be resummed \cite{Abel:2003ue}.  In this case one should also include a power of
$t^{\frac{3}{2}}$ in (\ref{eq:qfac_a2a3_finite}) coming from the resummation (\ref{eq:lim_fac_escl_resummed}), so there is no nett effect
upon $f\left(t\right)$.

As a check, one may wish to verify that the same power of $t$ is obtained when the term $A_{1}$ is factored onto the partition function.
It is then necessary to consider the term $\Phi\left(z_{1},z_{2},z_{3}\right)$ which is present in $A_{1}$ (displayed explicitly in figure
\ref{fig:correlators}).  Note that by (\ref{eq:extwist_f}), \begin{equation} \lim_{w\rightarrow
z}F_{3}\left(w,z\right)=1\label{eq:qfac_F}\end{equation} so that the first term in $\Phi$ becomes $\theta_{3}$. The second term in $\Phi$
vanishes by $\theta_{1}+\theta_{2}+\theta_{3}=1$, and hence one finds that \begin{equation} \lim_{z_{1}\rightarrow z_{2}\rightarrow
z_{3}}\sum_{i=1}^{3}\Phi^{i}\left(z_{1},z_{2},z_{3}\right)=m^{2}+1\,.\label{eq:qfac_Phi}\end{equation} Comparing (\ref{eq:quantum_a1}) and
(\ref{eq:quantum_a3}), we see that the rest of $A_{1}$ is of the same form as $A_{3}$ in the factorisation limit; therefore, the correct
power of $t$ will be obtained.

\subsection{The running coupling: logarithmic and power-law regimes\label{sub:running}}

At last, we have manufactured all the ingredients necessary to evaluate Yukawa coupling renormalization. We begin by considering the case
of an $\mathcal{N}=1$ supersymmetric set-up, where the amplitude comes purely from the term $A_{3}$, and again work with pinching limit 3
of figure \ref{fig:limits}.

It is now necessary to include the effects of the spin-dependent theta functions $\vartheta_{\alpha\beta}$, which is most easily done by
working with $A_{3}$ in the post-Riemann-identity form (\ref{eq:quantum_a3_afterriemann}).  Fixing $y_{1}=0$, making the replacement
(\ref{eq:qfac_thetarep}), using the result (\ref{eq:lim_untwist_detW_final}) for $\det W$ in the limit $y_{2}\rightarrow0$ and taking the
limit $y_{2}\rightarrow0$ in the full expression leads to the following explicit form for $A_{3}$:
\begin{multline}
A_{3}=-\overline{u}_{1}u_{2}\me^{-S_{\text{cl}}}k_{2}\cdot k_{3}y_{2}^{k_{1}\cdot k_{2}-\frac{3}{2}}\sinh\left(y_{3}\right)^{k_{1}\cdot k_{3}+\frac{1}{2}\left(\theta_{1}^{1}+\theta_{1}^{2}+\theta_{1}^{3}\right)-1}\sinh\left(y_{23}\right)^{k_{2}\cdot k_{3}+\frac{1}{2}\left(\theta_{2}^{1}+\theta_{2}^{2}+\theta_{2}^{3}\right)-3}\\
\left[\prod_{i=1}^{3}\left(\frac{\Gamma\left(1-\theta_{1}\right)\Gamma\left(1-\theta_{2}\right)\Gamma\left(1-\theta_{3}\right)}{4\pi}\right)^{\frac{1}{2}}\left(\left(2\pi y_{2}\right)^{1-\theta_{3}}\left(t-y_{3}\right)\right)^{-\frac{1}{2}}\sinh\left(\left(\theta_{3}^{i}-1\right)y_{3}\right)^{\frac{1}{2}}y_{2}^{\frac{1}{2}}\right]\\
\me^{-\pi t}\sinh\left(\left(1+\theta_{3}^{1}+\theta_{3}^{2}+\theta_{3}^{3}\right)y_{3}\right)\sinh\left(\left(1+\theta_{3}^{1}-\theta_{3}^{2}-\theta_{3}^{3}\right)y_{3}\right)\\
\sinh\left(\left(1-\theta_{3}^{1}+\theta_{3}^{2}-\theta_{3}^{3}\right)y_{3}\right)\sinh\left(\left(1-\theta_{3}^{1}-\theta_{3}^{2}+\theta_{3}^{3}\right)y_{3}\right)\,.\label{eq:run_a3}\end{multline}

To cast (\ref{eq:run_a3}) into a form that may be integrated, we note that it is dominated by large $y_{3},\, y_{23}$ and so we may
approximate the $\sinh$ functions as exponentials, leading to the rather more pleasant result\begin{equation}
A_{3}\simeq\me^{-S_{\text{cl}}}k_{2}\cdot k_{3}h\left(t\right)\me^{\left(k_{1}\cdot k_{3}+k_{2}\cdot k_{3}-m^{2}\right)\pi
y_{3}}\left(t-y_{3}\right)^{-\frac{3}{2}}y_{2}^{-1+k_{1}\cdot k_{2}+\frac{1}{2}m^{2}}\,,\label{eq:run_a3_2}\end{equation}
where\begin{equation} h\left(t\right)=\me^{-\pi t}\left(2\pi\right)^{\frac{1}{2}-k_{1}\cdot k_{3}-k_{2}\cdot
k_{3}+m^{2}}\prod_{i=1}^{3}\left[\frac{\Gamma\left(1-\theta_{1}^{i}\right)\Gamma\left(1-\theta_{2}^{i}\right)\Gamma\left(1-\theta_{3}^{i}\right)}{4\pi}\right]^{\frac{1}{2}}\label{eq:run_h}\end{equation}
with an overall phase dropped. The effect of $\me^{-S_{\text{cl}}}$ needs to be taken into account; specifically, one must decide whether
or not to perform the Poisson resummation (\ref{eq:lim_untwist_escl_resummed}).  The condition for the resummation to be valid is that
$2\pi\alpha^{\prime}\Delta t\gg L$, so that the integral depends upon the value of $\Delta t=t-y_{3}$; ergo, the result obtained will
depend upon the energy scale under consideration.

As $\Delta t\rightarrow\infty$, resummation of the Kaluza-Klein modes is appropriate. Then, the classical action
(\ref{eq:lim_untwist_escl_resummed}) conributes a factor\begin{equation}
\frac{\left(t-y_{3}\right)^{\frac{3}{2}}}{L_{1}L_{2}L_{3}}\label{eq:run_escl}\end{equation} to $A_{3}$ (we have set
$\alpha^{\prime}=\frac{1}{2}$ here). Choosing to make the change of variables (\ref{eq:qfac_cov}), the amplitude integral
reads\begin{align} \mathcal{A}_{3} &
=-\mi\int_{1/\Lambda^{2}}^{1/\mu^{2}}\frac{\mathrm{d}t}{t}f\left(t\right)t^{3}\int_{0}^{1}\sdif{\rho}\int_{1/t\Lambda^{\prime2}}^{1}\sdif{\lambda}A_{3}\label{eq:run_amp}\end{align}
At this point we should say something about the cut-offs on the integrals.  Firstly, note that we will require a UV cutoff $\Lambda$ on
our $t$-integral, despite the UV-finite nature of string theory. This cutoff is an artefact of making the large-$t$ approximation,
equivalent to sending $\alpha^{\prime}\rightarrow0$; if we were able to perform the appropriate elliptic integrals \emph{without} making
this approximation, we ought to find a UV-regular result. Secondly, we have placed a lower cut-off of $1/t\Lambda^{\prime2}$ on the
$\lambda$ integral. This may be viewed as a UV cutoff, since it removes the region where $y_{3}\rightarrow t$. With a suitable choice for
$\Lambda^{\prime}$, it also enforces the requirement that $2\pi\alpha^{\prime}\Delta t\gg L$ for the Kaluza-Klein resummation to be valid.

With our change of variables, we have\begin{multline}
\mathcal{A}_{3}=-\int_{1/\Lambda^{2}}^{1/\mu^{2}}\frac{\mathrm{d}t}{t}f\left(t\right)h\left(t\right)t^{2+k_{1}\cdot
k_{2}+\frac{1}{2}m^{2}} \left(L_{1}L_{2}L_{3}\right)^{-1}k_{2}\cdot k_{3}\\\int_{0}^{1}\sdif{\rho}\rho^{-1+k_{1}\cdot
k_{2}+\frac{1}{2}m^{2}}\int_{1/t\Lambda^{\prime2}}^{1}\sdif{\lambda}\me^{\left(k_{1}\cdot k_{3}+k_{2}\cdot k_{3}-m^{2}\right)\pi
t\left(1-\lambda\right)}\left(1-\lambda\right)^{k_{1}\cdot k_{2}+\frac{1}{2}m^{2}}\,.\label{eq:run_amp_log}\end{multline} As in the
previous section, the function of the $\rho$ integral is to supply a pole corresponding to the internal propagator in figure
\ref{fig:limits}. The $\lambda$ integral contains no poles when taken on-shell, so we may impose the on-shell conditions
(\ref{eq:qfac_onshell}) here \emph{before} performing the integral. Sending the cutoff $\Lambda^{\prime}\rightarrow\infty$,\begin{align}
\mathcal{A}_{3} & =-\left(L_{1}L_{2}L_{3}\right)^{-1}\frac{k_{2}\cdot k_{3}}{k_{1}\cdot
k_{2}+\frac{1}{2}m^{2}}\int_{1/\Lambda^{2}}^{1/\mu^{2}}\frac{\mathrm{d}t}{t}f\left(t\right)h\left(t\right)t^{2+k_{1}\cdot
k_{2}+\frac{1}{2}m^{2}}\nonumber \\ &
=\mathcal{Y}_{\text{qu}}\left(2\pi\right)^{-1}\left(L_{1}L_{2}L_{3}\right)^{-1}\int_{1/\Lambda^{2}}^{1/\mu^{2}}\frac{\mathrm{d}t}{t}\nonumber
\\ & \propto\log\left(\frac{\mu}{\Lambda}\right)\,.\label{eq:run_amp_log_eval}\end{align} Where in the second line we have set $m^{2}=0$
and cancelled the poles before taking everything on-shell. As one would expect, the string calculation reproduces the usual logarithmic
running of couplings at low energy.

At higher energy scales, $t$ can become sufficiently small that resummation of the Kaluza-Klein modes is no longer appropriate:
$2\pi\alpha^{\prime}\Delta t\ll L_{i}$ for all $L_{i}$. The classical action then contributes no powers of $\Delta t$ to the integral,
which now reads\begin{multline}
\mathcal{A}_{3}=-\int_{1/\Lambda^{2}}^{1/\mu^{2}}\frac{\mathrm{d}t}{t}f\left(t\right)h\left(t\right)t^{\frac{1}{2}+k_{1}\cdot
k_{2}+\frac{1}{2}m^{2}}\\ k_{2}\cdot k_{3}\int_{0}^{1}\sdif{\rho}\rho^{-1+k_{1}\cdot
k_{2}+\frac{1}{2}m^{2}}\int_{1/t\Lambda^{\prime2}}^{1}\sdif{\lambda}\me^{\left(k_{1}\cdot k_{3}+k_{2}\cdot k_{3}-m^{2}\right)\pi
t\left(1-\lambda\right)}\left(1-\lambda\right)^{k_{1}\cdot
k_{2}+\frac{1}{2}m^{2}}\lambda^{-\frac{3}{2}}\,.\label{eq:run_amp_power}\end{multline} Notice that this differs from
(\ref{eq:run_amp_log}) by a power of $t^{-\frac{3}{2}}$. Also, the $\lambda$ integral now has a pole as $\lambda\rightarrow0$
(i.e. $y_{3}\rightarrow t$), so that after performing the $\rho$ integral, cancelling poles and taking everything on-shell,

\begin{equation}
\mathcal{A}_{3}=\mathcal{Y}_{\text{qu}}\left(2\pi\right)^{-1}\int_{1/\Lambda^{2}}^{1/\mu^{2}}\sdif{t}t^{-\frac{5}{2}}\int_{1/t\Lambda^{\prime2}}^{1}\sdif{\lambda}\lambda^{-\frac{3}{2}}\,.\label{eq:run_amp_power_partial}\end{equation}
The IR and UV cutoffs in the two integrals should be associated with
each other, so that\begin{equation}
\mathcal{A}_{3}\propto\mu^{3}+\frac{3}{2}\Lambda^{2}\Lambda^{\prime}\,.\label{eq:run_amp_power_final}\end{equation}
We have obtained $\mu^{3}$ running, as one would expect for D$6$-branes
where three extra dimensions are present.

If the wrapping lengths $L_{i}$ are of mixed sizes, the amplitude may be an admixture of the two regimes, with the power of \textbf{$\mu$}
in (\ref{eq:run_amp_power_final}) dropping by one for each dimension which is not resummed. One may then envisage a situation where the
power-law behaviour changes as the energy scale $\mu$ increases, beginning first with logarithmic running and then switching to power-law
behaviour, with the power increasing as a greater number of extra dimensions open up (i.e. as $2\pi\alpha^{\prime}\Delta t$ exceeds the
brane wrapping lengths $L_{i}$ in the various tori).

In a non-supersymmetric model, the above results are modified in two ways. Firstly, the terms $A_{1}$ and $A_{2}$ no longer vanish via the
Riemann identity; therefore we have a vertex renormalization term plus an additional source of field renormalization. We will not evaluate
the vertex renormalization here (although we point out that sections \ref{sec:stringcalc}-\ref{sec:classical} contain all of the necessary
ingredients to do so), but instead examine how the field renormalization is modified. The function $f\left(t\right)$ which was determined
in sec. \ref{sub:qfac} is not modified in a nonsupersymmetric model; the only difference is that we can no longer set $m^{2}$ there, so
that the amplitudes $A_{2}$ and $A_{3}$ will now have a factor of $1-m^{2}$ on the bottom.

Evaluation of the term $A_{3}$ proceeds exactly as above. The cancellation of poles in \ref{eq:run_amp_log_eval} and
\ref{eq:run_amp_power_partial} appears now no longer valid, since we have $m^{2}\ne0$. However, including the term $A_{2}$ (which may be
calculated in an analogous manner to $A_{3}$, save for an intermediate expansion in $m^{2}$) leads to an extra contribution which
conspires to produce a cancellation of the same form as that in (\ref{eq:qfac_a2a3_poles}). This extra contribution cancels with the
factor of $1-m^{2}$ in $f\left(t\right)$, so that the field renormalization behaviour in non-supersymmetric models is the same as that in
supersymmetric models.

In the case where one of the branes goes through an orbifold fixed point, or we have O-planes in our model, the comments at the end of
section \ref{sec:quantum} apply -- the spin-dependent portions of the $A_{i}$ are modified. $f\left(t\right)$ should then be determined by
factoring onto a \emph{twisted} partition function, in which the modifications of the spin-dependent terms must be identical to those in
the $A_{i}$. Up to an overall normalization, then, $f\left(t\right)$ will be unchanged. The field renormalization may then be evaluated by
the procedure above, except that those portions of $A_{3}$ which come from spin-dependent terms (explicitly, the last two lines of
\ref{eq:run_a3}) must be appropriately modified.

\section{Summary and discussion\label{sec:summary}}

In this work, we have performed the first one-loop string calculation of Yukawa couplings on intersecting branes. We began by developing
the necessary technology for the calculation, in particular enumerating the selection rules on string states, describing the effects of
picture-changing on the vertex operators and developing open-string correlators for the spin and twist fields. We then applied this
technology to demonstrate the supersymmetric non-renormalization theorem, and identified the vertex and field renormalization
contributions to the amplitude.

Following this, we derived an expression for the classical instanton part of the twist fields, before examining them in the various
field-theory pinching limits of figure \ref{fig:limits}. In each limit, the classical action could be correctly factorized onto the
appropriate tree-level Yukawa coupling(s) and partition function piece. The origin of Kaluza-Klein modes in those limits which contain
bosons in the loop was identified.  Finally we showed how Yukawa beta functions in intersecting brane models may be extracted from the
calculation, finding that in the field-theory limit both logarithmic and power-law runnings are recovered depending upon the energy scale
under consideration.

One point that we did not address is the quantum normalization of the various limits. Note that the normalization which enters the
amplitude from the twist operators (via the $\left|\det W\right|^{-\frac{1}{2}}$ term) does not have the correct form (\ref{eq:qfac_y}),
and that this must be inserted into the normalization function $f\left(t\right)$ manually. Even so, one might expect that it ought to be
possible to recover the correct normalization factor by comparing the limits 2 and 3, which should contain factors of $\mathcal{Y}^{3}$
and $\mathcal{Y}$ respectively. Sadly, this is not the case: the problem is that limit 2 of figure \ref{fig:limits} contains only twisted
states in the loop, and thus ought to be factored onto a twisted partition function (a partition function between two branes at an
angle). Strictly speaking, the $f\left(t\right)$ obtained is now \emph{different}. It is then be necessary to include manually a
$\mathcal{Y}^{3}$ factor, and our reasoning becomes circular. We conclude that our three-point calculation cannot be used to extract the
quantum normalization factor (\ref{eq:qfac_y}) -- this is exactly the situation at tree-level, where the normalization must be obtained by
factorization of a four-point correlator onto a three-point correlator
\cite{Burwick:1990tu,Stieberger:1992bj,Erler:1992gt,Stieberger:1992vb,Bailin:1993km,Cvetic:2003ch,Abel:2003vv,Lust:2004cx}.

As a final remark, we note that much of the conformal field theory framework developed in this paper is applicable to one-loop
calculations on intersecting branes in general (and also in fact to orbifold models, providing an alternative to the calculations
presented in \cite{Minahan:1987ha}). In particular it would be interesting to apply our results to a four-point correlator at one-loop,
which contains information about the one-loop Kähler potential \cite{Lust:2004cx}.  Furthermore, factorization of this four-point
amplitude onto a three-point amplitude should explicitly reproduce the Yukawa normalization factor (\ref{eq:qfac_y}).

\acknowledgments{It is a pleasure to thank Mirjam Cveti\v{c}, Dumitru Ghilencea and Mark Goodsell for helpful discussions. This work was
partially supported by a PPARC studentship.}

\appendix

\section{The Jacobi $\vartheta$-functions\label{sec:theta}}

We take the usual definition for the Jacobi $\vartheta$-functions \cite{Mumford:1982,Polchinski:1998rq},\begin{align}
\vartheta_{00}\left(z,\mi t\right)= & \sum_{n=-\infty}^{\infty}\me^{-\pi n^{2}t}\me^{2\mi\pi nz}\nonumber \\ = & 1+2\me^{-\pi
t}\cos\left(2\pi z\right)+2\me^{-4\pi t}\cos\left(4\pi z\right)+\cdots\label{eq:theta_00}\\ \vartheta_{01}\left(z,\mi t\right)= &
\sum_{n=-\infty}^{\infty}\left(-1\right)^{n}\me^{-\pi n^{2}t}\me^{2\mi\pi nz}\nonumber \\ = & 1-2\me^{-\pi t}\cos\left(2\pi
z\right)+2\me^{-4\pi t}\cos\left(4\pi z\right)-\cdots\label{eq:theta_01}\\ \vartheta_{10}\left(z,\mi t\right)= &
\sum_{n=-\infty}^{\infty}\me^{-\pi\left(n-\frac{1}{2}\right)^{2}t}\me^{2\mi\pi\left(n-\frac{1}{2}\right)z}\nonumber \\ = & 2\me^{-\pi
t/4}\cos\left(\pi z\right)+2\me^{-9\pi t/4}\cos\left(3\pi z\right)+\cdots\label{eq:theta_10}\\ \vartheta_{11}\left(z,\mi t\right)= &
\sum_{n=-\infty}^{\infty}\left(-1\right)^{n-\frac{1}{2}}\me^{-\pi\left(n-\frac{1}{2}\right)^{2}t}\me^{2\mi\pi\left(n-\frac{1}{2}\right)z}\nonumber
\\ = & -2\me^{-\pi t/4}\sin\left(\pi z\right)+2\me^{-9\pi t/4}\sin\left(3\pi z\right)+\cdots\label{eq:theta_11}\end{align} We also define
$\vartheta_{1}\left(z|\mi t\right)=-\vartheta_{11}\left(z,\mi t\right)$.

The $\vartheta$-functions obey a useful `Riemann' identity,\begin{equation}
\sum_{\alpha\beta}\delta_{\alpha\beta}\vartheta_{\alpha\beta}\left(x\right)\vartheta_{\alpha\beta}\left(y\right)\vartheta_{\alpha\beta}\left(u\right)\vartheta_{\alpha\beta}\left(v\right)=2\vartheta_{1}\left(x^{\prime}\right)\vartheta_{1}\left(y^{\prime}\right)\vartheta_{1}\left(u^{\prime}\right)\vartheta_{1}\left(v^{\prime}\right)\,,\label{eq:riemann}\end{equation}
where $\delta_{00}=\delta_{11}=+1$, $\delta_{01}=\delta_{10}=-1$ and\begin{eqnarray} x^{\prime}=\tfrac{1}{2}\left(x+y+u+v\right) & \qquad
& u^{\prime}=\tfrac{1}{2}\left(x+y-u-v\right)\nonumber \\ y^{\prime}=\tfrac{1}{2}\left(x-y+u-v\right) & \qquad &
v^{\prime}=\tfrac{1}{2}\left(x-y-u+v\right)\,.\label{eq:riemann_rels}\end{eqnarray}

\section{W integrals in various limits\label{sec:Wlimits}}

Information about the geometry of the problem is contained in the $W$-integrals (\ref{eq:twist_wigrls}), which we examine in the
$t\rightarrow\infty$ limit. Begin with integrals around $\gamma_{1}$, which we take to be along the imaginary axis $z=\mi q$. Using the
expansion (\ref{eq:theta_11}) together with the result $\theta_{1}+\theta_{2}+\theta_{3}=1$, one can write the integrals up to
exponentially suppressed terms as\begin{multline}
W_{1}^{1}=\mi\int_{\left(y_{3}-t\right)/2}^{\left(y_{3}+t\right)/2}\sdif{q}\left(\mi\sinh\pi
q\right)^{\theta_{1}-1}\left(\mi\sinh\pi\left(q-y_{2}\right)\right)^{\theta_{2}}\left(\mi\sinh\pi\left(q-y_{3}\right)\right)^{\theta_{3}-1}\\
\mi\sinh\pi\left(q+\theta_{2}y_{2}+\left(\theta_{3}-1\right)y_{3}\right)\end{multline}
\begin{multline}
W_{1}^{2}=\mi\int_{\left(y_{3}-t\right)/2}^{\left(y_{3}+t\right)/2}\sdif{q}\left(\mi\sinh\pi q\right)^{\theta_{1}}\left(\mi\sinh\pi\left(q-y_{2}\right)\right)^{\theta_{2}-1}\left(\mi\sinh\pi\left(q-y_{3}\right)\right)^{\theta_{3}-1}\\
\mi\sinh\pi\left(q+\left(\theta_{2}-1\right)y_{2}+\left(\theta_{3}-1\right)y_{3}\right)\end{multline}
\begin{multline}
W_{1}^{3}=-\mi\int_{-\left(y_{3}+t\right)/2}^{-\left(y_{3}-t\right)/2}\sdif{q}\overline{\left(-\mi\sinh\pi
q\right)^{-\theta_{1}}\left(-\mi\sinh\pi\left(q+y_{2}\right)\right)^{-\theta_{2}}\left(-\mi\sinh\pi\left(q+y_{3}\right)\right)^{-\theta_{3}}}\\
\overline{-\mi\sinh\pi\left(q+\theta_{2}y_{2}+\theta_{3}y_{3}\right)}\,,\label{eq:wigrls_larget}\end{multline} where the translational
invariance of the problem has been used to set $y_{1}=0$ (in what follows we always assume $0<y_{1}<y_{2}<t$).  Our approximation has lost
the periodicity of $\vartheta_{1}$ in the imaginary direction; therefore, we have chosen limits on the integrals so as to keep the branch
cuts in the centre of the worldsheet, ensuring that we integrate over one complete period.

Integrals around around $\gamma_{2}$ are simpler. Using the same approximation, one finds integrals of sine functions rather than sinh
functions, taken along the real axis. In this case, we always find $W_{2}^{i}=1$. Finally, integrals around $\gamma_{3}$ are of the same
form as those around $\gamma_{1}$ but with the limits taken as $0$ and $y_{2}$, and the overall result multiplied by a phase
factor\begin{equation}
P=4\me^{\mi\pi\left(\theta_{2}-\theta_{1}\right)}\sin\left(\pi\theta_{2}\right)\sin\left(\pi\theta_{1}\right)\,,\label{eq:pochfactor}\end{equation}
which comes from the Pochhammer loop \cite{Abel:2003vv}.

\subsection{Limit 0: $t\rightarrow\infty$, $y_{1}\rightarrow y_{2}\rightarrow y_{3}$\label{sub:Wlimit0}}

In this limit, we may expand (\ref{eq:wigrls_larget}) in $y_{1}$ and $y_{2}$ under the integrals, with the results\begin{equation}
W_{1}^{1}=W_{1}^{2}=\mi t+\mathcal{O}\left(y_{i}\right)\qquad\qquad W_{1}^{3}=-\mi
t+\mathcal{O}\left(y_{i}\right)\,,\label{eq:lim_fac_w_gamma1}\end{equation} which give $W$ the form\begin{equation}
W\simeq\left(\begin{array}{ccc} \mi t & \mi t & -\mi t\\ 1 & 1 & 1\\ W_{3}^{1} & W_{3}^{2} &
W_{3}^{3}\end{array}\right)\label{eq:lim_fac_wmat}\end{equation} and hence\begin{equation} \left|\det
W\right|=2t\left(W_{3}^{1}-W_{3}^{2}\right)\,.\label{eq:lim_fac_detW}\end{equation} Trigonometric identities simplify this combination
to\begin{multline} W_{3}^{1}-W_{3}^{2}=\mi P\sinh\left(\pi
y_{2}\right)\sinh\left(\pi\left(y_{2}\theta_{2}+y_{3}\left(\theta_{3}-1\right)\right)\right)\\ \int_{0}^{y_{2}}\sdif{q}\left(\mi\sinh\pi
q\right)^{\theta_{1}-1}\left(\mi\sinh\pi\left(q-y_{2}\right)\right)^{\theta_{2}-1}\left(\mi\sinh\pi\left(q-y_{3}\right)\right)^{\theta_{3}-1}\,,\label{eq:lim_fac_combo}\end{multline}
and in the limit of small $y_{i}$ the integral reduces to one which may be evaluated using the Gauss hypergeometric function
$_{2}F_{1}$,\begin{multline}
\int_{0}^{y_{2}}\sdif{q}q^{\theta_{1}-1}\left(q-y_{2}\right)^{\theta_{2}-1}\left(q-y_{3}\right)^{\theta_{3}-1}\\
=\frac{\Gamma\left(\theta_{1}\right)\Gamma\left(\theta_{2}\right)}{\Gamma\left(\theta_{1}+\theta_{2}\right)}{}_{2}F_{1}\left(\theta_{1},1-\theta_{3},\theta_{1}+\theta_{2};\frac{y_{2}}{y_{3}}\right)y_{2}^{\theta_{1}+\theta_{2}-1}y_{3}^{\theta_{3}-1}\,.\label{eq:lim_fac_h2f1_int}\end{multline}
Including the phase factor $P$ and using the condition $\theta_{1}+\theta_{2}+\theta_{3}=1$, plus the identity
$_{2}F_{1}\left(a,b,b;z\right)=\left(1-z\right)^{-a}$, we end up with\begin{equation} \left|\det
W\right|=\frac{8\pi^{2}t}{\Gamma\left(1-\theta_{1}\right)\Gamma\left(1-\theta_{2}\right)\Gamma\left(1-\theta_{3}\right)}y_{2}^{-\theta_{3}}y_{3}^{-\theta_{2}}\left(y_{3}-y_{2}\right)^{-\theta_{1}}y_{2}\left(\theta_{2}y_{2}-\left(1-\theta_{3}\right)y_{3}\right)\,.\label{eq:lim_fac_detW_evaluated}\end{equation}
If desired, we may reinstate $y_{1}$ by requiring that $\det W$ be translationally invariant; the result is shown in equation
(\ref{eq:qfac_detW}).  We will also make use of $W_{3}^{3}$, which may be evaluated through use of hypergeometric functions
as\begin{equation}
W_{3}^{3}=\frac{4\pi^{2}y_{12}^{\theta_{3}}y_{13}^{\theta_{2}}y_{23}^{\theta_{1}}}{\Gamma(1-\theta_{1})\Gamma(1-\theta_{2})\Gamma(1-\theta_{3})}\,.\label{eq:lim_fac_W33}\end{equation}

\subsection{Limit 1: $\left(t-y_{3}\right)\rightarrow\infty$, generic $y_{i}$\label{sub:Wlimit1}}

\FIGURE[t]{\scalebox{0.9}{\includegraphics{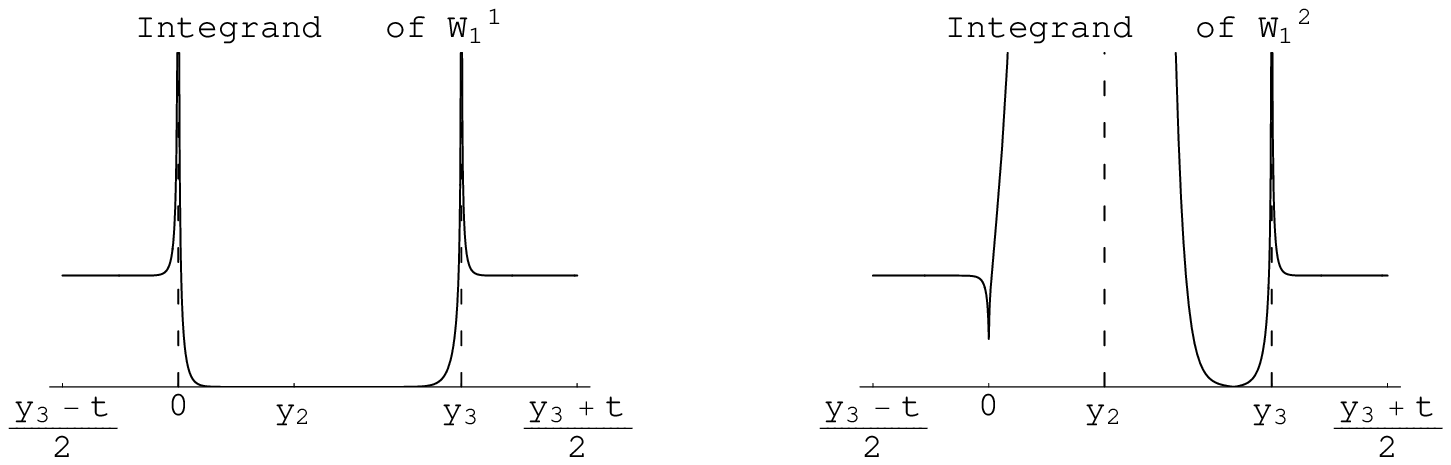}}
\caption{Absolute value of the $W_{1}^{i}$ integrands, for arbitrarily chosen angles $0<\theta_{j}<\frac{1}{2}$ and generic $y_{i}$. Not
shown is $W_{1}^{3}$, which is simply a reflection of $W_{1}^{1}$ in the $y$-axis.\label{fig:plots}} } The structure of our integrands in
this case is shown in figure \ref{fig:plots}.  The integrands of $W_{1}^{1}$ and $W_{3}^{1}$ both contain poles at $q=0$ and $q=y_{3}$,
but are regular elsewhere. Let us first consider integrals about $\gamma_{1}$. One may expand the $W_{1}^{1}$ integrand about the $q=0$
pole as follows:\begin{align} \left.W_{1}^{1}\right|_{0} &
=\mi\left(-\frac{1}{2}\right)^{1-\theta_{1}}\int_{\left(y_{3}-t\right)/2}^{\infty}\sdif{q}\left(\sinh\pi q\right)^{\theta_{1}-1}e^{\pi
q\left(\theta_{1}-1\right)}\nonumber \\ &
=\mi\frac{\left(-1\right)^{1-\theta_{1}}}{2\pi}B\left(\me^{\pi\left(t-y_{3}\right)};1-\theta_{1},\theta_{1}\right)+1\,,\label{eq:lim_gen_w1pole1}\end{align}
where $B\left(z;a,b\right)$ is the incomplete Euler Beta function.  Expanding around the $q=y_{3}$ pole gives a similar result,
\begin{align} \left.W_{1}^{1}\right|_{y_{3}} &
=\mi\left(\frac{1}{2}\right)^{1-\theta_{3}}\int_{-\infty}^{\left(y_{3}+t\right)/2}\sdif{q}\left(\sinh\pi\left(q-y_{3}\right)\right)^{\theta_{3}-1}e^{\pi\left(q-y_{3}\right)\left(\theta_{3}-1\right)}\nonumber
\\ & =\mi\left(\frac{1}{2}\right)^{1-\theta_{3}}\int_{-\infty}^{\left(t-y_{3}\right)/2}\sdif{q^{\prime}}\left(\sinh\pi
q^{\prime}\right)^{\theta_{3}-1}e^{\pi q^{\prime}\left(\theta_{3}-1\right)}\nonumber \\ &
=\mi\frac{\left(-1\right)^{1-\theta_{3}}}{2\pi}B\left(\me^{\pi\left(t-y_{3}\right)};1-\theta_{3},\theta_{3}\right)\,.\label{eq:lim_gen_w1pole2}\end{align}
The full $W_{1}^{1}$ is then found as the sum of these two,\begin{equation}
W_{1}^{1}=\frac{\mi}{2\pi}\left[\left(-1\right)^{1-\theta_{1}}B\left(\me^{\pi\left(t-y_{3}\right)};1-\theta_{1},\theta_{1}\right)+\left(-1\right)^{1-\theta_{3}}B\left(\me^{\pi\left(t-y_{3}\right)};1-\theta_{3},\theta_{3}\right)\right]+1\,.\label{eq:lim_gen_w11}\end{equation}
Since we are interested in the limit where $t-y_{3}\rightarrow\infty$, we may apply the large-$z$ expansion of the incomplete Euler Beta
function,\begin{equation} B(z;a,b)=\left(-1\right)^{-a}\left(\mi\pi+\log
z-\psi\left(a\right)-\gamma_{E}\right)+\mathcal{O}\left(z^{-1}\right)\qquad\left(\left|z\right|\rightarrow\infty,\,
a+b=1\right),\label{eq:beta_largez}\end{equation} with the result\begin{equation}
W_{1}^{1}=\mi\left(t-y_{3}\right)-\frac{\mi}{2\pi}\left(2\gamma_{E}+\psi\left(1-\theta_{1}\right)+\psi\left(1-\theta_{3}\right)\right)+\mathcal{O}\left(t^{-1}\right)\,.\label{eq:lim_gen_w11_final}\end{equation}

Now consider the $W_{1}^{3}$ integral, which is similar. An analogous expansion about the poles at $q=0$ and $q=-y_{3}$ leads
to\begin{equation}
W_{1}^{3}=-\frac{\mi}{2\pi}\left[\left(-1\right)^{\theta_{1}}B\left(\me^{\pi\left(t-y_{3}\right)};\theta_{1},1-\theta_{1}\right)+\left(-1\right)^{\theta_{3}}B\left(\me^{\pi\left(t-y_{3}\right)};\theta_{3},1-\theta_{3}\right)\right]-1\label{eq:lim_gen_w13}\end{equation}
and using the large-$z$ expansion (\ref{eq:beta_largez}),\begin{equation}
W_{1}^{3}=-\mi\left(t-y_{3}\right)+\frac{\mi}{2\pi}\left(2\gamma_{E}+\psi\left(\theta_{1}\right)+\psi\left(\theta_{3}\right)\right)+\mathcal{O}\left(t^{-1}\right)\,.\label{eq:lim_gen_w13_final}\end{equation}
The integrands for $W_{1}^{2}$ have poles at $q=y_{2}$ and $q=y_{3}$; however, the pole at $y_{2}$ is exponentially dominant for generic
$y_{2}$. Therefore, the leading term in $W_{1}^{2}$ for generic $y_{2}$ is \begin{align} \left.W_{1}^{2}\right|_{y_{2}} &
=\mi\left(-1\right)^{\theta_{3}}2^{\theta_{2}-1}\me^{\pi
y_{2}\left(1-\theta_{2}\right)}\int_{-\infty}^{\infty}\sdif{q}\left(\sinh\pi\left(q-y_{2}\right)\right)^{\theta_{2}-1}\me^{\pi
q\left(\theta_{1}-\theta_{3}\right)}\nonumber \\ & =\frac{\mi\me^{2\pi
y_{2}\theta_{1}}\Gamma\left(\theta_{1}\right)\Gamma\left(\theta_{3}\right)}{2\pi\Gamma\left(1-\theta_{2}\right)}\,.\label{eq:lim_gen_w12}\end{align}
We could also extract the sub-leading behaviour, which we expect to be of a similar form to the result in $W_{1}^{1}$, but as we will see
below it is unnecessary for our calculation.

Integrals around the Pochhammer cycle $\gamma_{3}$ may be found using the same technique of expanding about poles in $q$, but keeping only
the portion of the integral which corresponds to the region $0<q<y_{2}$.  For instance, $W_{3}^{1}$ is found from the positive portion of
the expansion around the first pole multiplied by the Pochhammer factor (\ref{eq:pochfactor}),\begin{align} W_{3}^{1} & =\mi
P\left.W_{1}^{1}\right|_{0,q>0}\nonumber \\ & =-\mi P\frac{(-1)^{\theta_{1}}}{2\,\sin(\pi\theta_{1})}\nonumber \\ &
=-2\mi\me^{\mi\pi\theta_{2}}\sin\left(\pi\theta_{2}\right)\,.\label{eq:lim_gen_w31}\end{align} (The integrals $W_{1}^{i}$ are along the
principal branch, and so $(-1)=\me^{\mi\pi}$.) Repeating the procedure for $W_{3}^{3}$ leads to an identical result,\begin{equation}
W_{3}^{3}=-2\mi\me^{\mi\pi\theta_{2}}\sin\left(\pi\theta_{2}\right)\,.\label{eq:lim_gen_w33}\end{equation} For generic $y_{2}$, the
leading term in the $W_{3}^{2}$ integral is the portion of the$\left.W_{1}^{2}\right|_{y_{2}}$ integral which is below \emph{}the pole at
$q=y_{2}$, multiplied by the Pochhammer factor,\begin{align} W_{3}^{2} & \simeq\mi P\left.W_{1}^{2}\right|_{y_{2},q<y_{2}}\nonumber \\ &
=\mi P\frac{\left(-1\right)^{\theta_{1}}\me^{2\pi
y_{2}\theta_{1}}\Gamma\left(\theta_{1}\right)\Gamma\left(\theta_{2}\right)}{2\pi\Gamma\left(1-\theta_{3}\right)}\nonumber \\ &
=\frac{2\mi\pi\me^{\mi\pi\theta_{2}}\me^{2\pi
y_{2}\theta_{1}}}{\Gamma\left(1-\theta_{1}\right)\Gamma\left(1-\theta_{2}\right)\Gamma\left(1-\theta_{3}\right)}\,.\label{eq:lim_gen_w32}\end{align}
Going from the second line to the third makes use of the reflection identity
$\Gamma\left(\theta\right)\Gamma\left(1-\theta\right)=\pi\csc\left(\pi\theta\right)$.

It proves convenient to define a set of functions as follows,\begin{eqnarray} A & = &
-\frac{\mi}{2}\left[\cot\left(\pi\theta_{1}\right)+\cot\left(\pi\theta_{3}\right)\right]\,=\,-\frac{i}{2}\left(\frac{\sin(\pi\theta_{2})}{\sin(\pi\theta_{1})\sin(\pi\theta_{3})}\right)\nonumber
\\ B & = & -2\mi e^{i\pi\theta_{2}}\sin(\pi\theta_{2})\nonumber \\ C & = &
-\frac{\Gamma(\theta_{2})}{\Gamma(1-\theta_{1})\Gamma(1-\theta_{3})}e^{2\pi y_{2}\theta_{1}}\,,\nonumber \\ D & = &
-\frac{\mi}{2\pi}\left[2\gamma_{E}+\psi\left(1-\theta_{1}\right)+\psi\left(1-\theta_{3}\right)\right]\label{eq:lim_gen_funcs}\end{eqnarray}
in terms of which the $W$ matrix may be expressed as\begin{equation} W=\left(\begin{array}{ccc} \mi\Delta t+D & AC & -\mi\Delta t+D+A\\ 1
& 1 & 1\\ B & BC & B\end{array}\right)\,,\label{eq:lim_gen_wmat}\end{equation} where $\Delta t=t-y_{3}$, we have dropped terms of order
$t^{-1}$ and also used the relation\[ \psi\left(\theta\right)=\psi\left(1-\theta\right)-\pi\cot\left(\pi\theta\right)\,.\] When we take
the determinant, the dependence on $W_{1}^{2}=AC$ drops out, leaving\begin{align} \left|\det W\right| &
=\left|B\left(1-C\right)\left(A+2D-2\mi\Delta t\right)\right|\label{eq:lim_gen_detW_tmp}\\ & \simeq2\left|BC\right|\Delta t\,.\nonumber
\end{align} Explicitly, the leading term is\begin{equation} \left|\det W\right|\simeq\frac{4\me^{2\pi
y_{2}\theta_{1}}\left(t-y_{3}\right)}{\Gamma\left(1-\theta_{1}\right)\Gamma\left(1-\theta_{2}\right)\Gamma\left(1-\theta_{3}\right)}\,.\label{eq:lim_gen_detW}\end{equation}
The angular factor in front of this expression is identical to that found in (\ref{eq:lim_fac_detW_evaluated}).

\subsection{Limit 2: $\left(t-y_{3}\right)\rightarrow0$, generic $y_{i}$\label{sub:Wlimit2}}

As far as the $W$-integrals are concerned, the difference between this case and the first lies in only the integrals about $\gamma_{1}$.
The expressions (\ref{eq:lim_gen_w11}) and (\ref{eq:lim_gen_w13}) for $W_{1}^{1}$ and $W_{1}^{3}$ are still valid in this limit, and their
final form is obtained by using the expansion of the incomplete Euler beta function about $z=1$,\begin{equation}
B(z;a,b)=B\left(a,b\right)+\mathcal{O}\left(z^{a}\left(1-z\right)^{b}\right)\,,\label{eq:beta_smallz}\end{equation} with
$B\left(a,b\right)=\Gamma\left(a\right)\Gamma\left(b\right)/\Gamma\left(a+b\right)$ the ordinary Euler beta function. We
find\begin{multline}
W_{1}^{1}=-\frac{\mi}{2}\left[\cot\left(\pi\theta_{1}\right)+\cot\left(\pi\theta_{3}\right)\right]+\frac{\mi}{2}\left[\frac{\pi^{\theta_{1}-1}}{\theta_{1}}\left(t-y_{3}\right)^{\theta_{1}}+\frac{\pi^{\theta_{3}-1}}{\theta_{3}}\left(t-y_{3}\right)^{\theta_{3}}\right]\\
+\mathcal{O}\left(\left(t-y_{3}\right)^{1+\theta_{i}}\right)\,,\label{eq:lim_twist_w11}\end{multline} and,\begin{multline}
W_{1}^{3}=-\frac{\mi}{2}\left[\cot\left(\pi\theta_{1}\right)+\cot\left(\pi\theta_{3}\right)\right]-\frac{\mi}{2}\left[\frac{\pi^{-\theta_{1}}}{1-\theta_{1}}\left(t-y_{3}\right)^{1-\theta_{1}}+\frac{\pi^{-\theta_{3}}}{1-\theta_{3}}\left(t-y_{3}\right)^{1-\theta_{3}}\right]\\
+\mathcal{O}\left(\left(t-y_{3}\right)^{2-\theta_{i}}\right)\,.\label{eq:lim_twist_w13}\end{multline}

The integral $W_{1}^{2}$ is unchanged from before, as are all integrals around $\gamma_{2}$ and $\gamma_{3}$. Writing\begin{eqnarray}
\Delta t_{1}^{\prime} & = &
\frac{\mi}{2}\left[\frac{\pi^{\theta_{1}-1}}{\theta_{1}}\left(t-y_{3}\right)^{\theta_{1}}+\frac{\pi^{\theta_{3}-1}}{\theta_{3}}\left(t-y_{3}\right)^{\theta_{3}}\right]\nonumber
\\ \Delta t_{3}^{\prime} & = &
\frac{\mi}{2}\left[\frac{\pi^{-\theta_{1}}}{1-\theta_{1}}\left(t-y_{3}\right)^{1-\theta_{1}}+\frac{\pi^{-\theta_{3}}}{1-\theta_{3}}\left(t-y_{3}\right)^{1-\theta_{3}}\right]\,,\label{eq:lim_twist_deltats}\end{eqnarray}
and using the definitions (\ref{eq:lim_gen_funcs}), the $W$ matrix is\begin{equation} W=\left(\begin{array}{ccc} A+i\Delta t_{1}^{\prime}
& AC & A-i\Delta t_{3}^{\prime}\\ 1 & 1 & 1\\ B & BC & B\end{array}\right)\,.\label{eq:lim_twist_wmat}\end{equation}

Note that the $t$-independent terms in $W_{1}^{1}$ and $W_{1}^{3}$ now have the \emph{same} sign, whereas in the previous section their
signs were different. This means that they drop out of the determinant to leave
\begin{equation}
\left|\det W\right|\simeq\frac{4\me^{2\pi y_{2}\theta_{1}}\left(\Delta t_{1}^{\prime}+\Delta t_{3}^{\prime}\right)}{\Gamma\left(1-\theta_{1}\right)\Gamma\left(1-\theta_{2}\right)\Gamma\left(1-\theta_{3}\right)}\,.\label{eq:lim_twist_wmat_final}\end{equation}

\subsection{Limit 3: $\left(t-y_{3}\right)\rightarrow\infty$, $y_{2}\rightarrow y_{1}$\label{sub:Wlimit3}}

In this limit, an appropriate approach is to expand in $y_{2}$ under each integral and keep the leading term. For $W_{1}^{1}$, this gives
\begin{equation}
W_{1}^{1}\simeq\mi\int_{\left(y_{3}-t\right)/2}^{\left(y_{3}+t\right)/2}\sdif{q}\left(\mi\sinh\pi
q\right)^{-\theta_{3}}\left(\mi\sinh\pi\left(q-y_{3}\right)\right)^{\theta_{3}-1}\mi\sinh\pi\left(q+\left(\theta_{3}-1\right)y_{3}\right)\,,\label{eq:lim_untwist_W11igrl}\end{equation}
which has poles at $q=0$ and $q=y_{3}$. As before, one may expand about these poles, integrate and use the expansion
(\ref{eq:beta_largez}), leading to\begin{equation}
W_{1}^{1}=\mi\left(t-y_{3}\right)-\frac{\mi}{2\pi}\left(2\gamma_{E}+\psi\left(\theta_{3}\right)+\psi\left(1-\theta_{3}\right)\right)+\mathcal{O}\left(t^{-1}\right)\,.\label{eq:lim_untwist_W11}\end{equation}
Other integrals about $\gamma_{1}$ are found in a similar fashion, with the result
$W_{1}^{1}=W_{1}^{2}=-W_{1}^{3}$. Accordingly,\begin{equation} W=\left(\begin{array}{ccc} W_{1}^{1} & W_{1}^{1} & -W_{1}^{1}\\ 1 & 1 & 1\\
W_{3}^{1} & W_{3}^{2} & W_{3}^{3}\end{array}\right)\,,\label{eq:lim_untwist_Wmat}\end{equation} and so\begin{equation} \left|\det
W\right|=2W_{1}^{1}\left(W_{3}^{1}-W_{3}^{2}\right)\,.\label{eq:lim_untwist_detW}\end{equation}

The integrals about $\gamma_{3}$ may performed by approximating them as, for instance,\begin{align} W_{3}^{1} & =\mi
P\left(-1\right)^{\theta_{1}-1}\left(2\pi\right)^{-\theta_{3}}\int_{0}^{y_{2}}\sdif{q}\me^{-\pi\theta_{3}\left(y_{2}-q\right)}q^{\theta_{1}-1}\left(y_{2}-q\right)^{\theta_{2}}\nonumber
\\ & =\mi
P\left(-1\right)^{\theta_{1}-1}\left(-2\pi\right)^{-\theta_{3}}y_{2}^{1-\theta_{3}}\me^{-\pi\theta_{3}y_{2}}\int_{0}^{1}\sdif{\lambda}\me^{\pi\theta_{3}y_{2}\lambda}\lambda^{\theta_{1}-1}\left(1-\lambda\right)^{\theta_{2}}\,,\label{eq:lim_untwist_w31approx}\end{align}
and using the integral\begin{equation}
\int_{0}^{1}\sdif{\lambda}\me^{z\lambda}\lambda^{a-1}\left(1-\lambda\right)^{-a+b-1}=\frac{\Gamma\left(a\right)\Gamma\left(b-a\right)}{\Gamma\left(b\right)}+\mathcal{O}\left(z\right)\,,\label{eq:lim_untwist_kummer}\end{equation}
which may be found via the Kummer hypergeometric function ${}_{1}F_{1}$.  The final results are
\begin{eqnarray}
W_{3}^{1} & = &
\frac{2\pi\mi\me^{\mi\pi\theta_{2}}}{\Gamma\left(1-\theta_{1}\right)\Gamma\left(1-\theta_{2}\right)\Gamma\left(1-\theta_{3}\right)}\left(\frac{\theta_{2}}{1-\theta_{3}}\right)\left(2\pi
y_{2}\right)^{1-\theta_{3}}\nonumber \\ W_{3}^{2} & = &
-\frac{2\pi\mi\me^{\mi\pi\theta_{2}}}{\Gamma\left(1-\theta_{1}\right)\Gamma\left(1-\theta_{2}\right)\Gamma\left(1-\theta_{3}\right)}\left(\frac{\theta_{1}}{1-\theta_{3}}\right)\left(2\pi
y_{2}\right)^{1-\theta_{3}}\nonumber \\ W_{3}^{3} & = &
\frac{2\pi\mi\me^{\mi\pi\left(\theta_{2}-2\theta_{1}\right)}}{\Gamma\left(\theta_{1}\right)\Gamma\left(\theta_{2}\right)\Gamma\left(\theta_{3}\right)}\left(\frac{1}{\theta_{3}}\right)\left(2\pi
y_{2}\right)^{\theta_{3}}\,,\label{eq:lim_untwist_W3i}\end{eqnarray} and the determinant therefore has the leading term\begin{equation}
\left|\det W\right|\simeq\frac{4\pi\left(2\pi
y_{2}\right)^{1-\theta_{3}}}{\Gamma\left(1-\theta_{1}\right)\Gamma\left(1-\theta_{2}\right)\Gamma\left(1-\theta_{3}\right)}\left(t-y_{3}\right)\,.\label{eq:lim_untwist_detW_final}\end{equation}

\bibliographystyle{JHEP}
\bibliography{yukawa}

\end{document}